\documentclass[prb,aps,twocolumn,amsmath,amssymb,floatfix,superscriptaddress]{revtex4}

\usepackage[dvips]{graphicx}
\usepackage{xcolor}
\usepackage{bbold}
\usepackage{physics}
\usepackage{soul}
\definecolor{dred}{rgb}{0.5,0.2,0}
\usepackage[colorlinks=true, citecolor=blue, urlcolor=blue]{hyperref}
\usepackage{tabularx}
\usepackage{array}[=2016-10-06]

\date{\today}

\begin{document}

\title{Fibonacci-Engineered Spin and Charge 
Thermoelectrics in a Long Range Su-Schrieffer-Heeger Chain: A Pathway to Giant Figure of Merit}

\author{Ranjini Bhattacharya\footnote{Corresponding author: \href{mailto:ranjinibhattacharya@gmail.com}{ranjinibhattacharya@gmail.com}}}

\email{ranjinibhattacharya@gmail.com}

\affiliation{Department of Condensed Matter and Materials Physics, S. N. Bose National Centre for Basic Sciences,
JD-Block, Sector III, Salt Lake, Kolkata 700098, India}

\author{Souvik Roy}

\email{souvikroy138@gmail.com}

\affiliation{School of Physical Sciences, National Institute of Science Education and Research, Jatni 752050, India}
\affiliation{Homi Bhabha National Institute, Training School Complex, Anushaktinagar, Mumbai 400094, India}

\begin{abstract}
In this work, we present a \textcolor{black}{novel} investigation into the spin-dependent thermoelectric performance of an extended Su–Schrieffer–Heeger (SSH) model, showcasing for the first time how its intrinsic spin filtration mechanism can be strategically harnessed to function as an efficient spin thermoelectric generator. By introducing a Fibonacci-type aperiodic modulation in the onsite energies, we engineer a deterministic disorder that mimics realistic aperiodic systems and profoundly influences transport characteristics. Furthermore, we incorporate both nearest-neighbor (NN) and next-nearest-neighbor (NNN) hopping amplitudes with tunable cosine dependencies, enabling us to meticulously explore the intricate interplay between these hopping processes and its implications on thermoelectric behavior. Our analysis reveals a remarkable  enhancement in the dimensionless thermoelectric figure of merit $ZT$ for both charge and spin transport channels, under carefully optimized conditions. Notably, the spin thermoelectric response exhibits distinct advantages, opening a new frontier in the design of next-generation thermoelectric materials and devices. This \textcolor{black}{qualitative} study not only \textcolor{black}{deepens} our understanding of aperiodic topological systems but also establish a foundational framework for exploiting spin-based thermoelectricity in low-dimensional platforms.
\end{abstract}

\maketitle

\section{\label{sec:intro}Introduction}

With the ever-increasing global energy demand, compounded by the pressing challenges of climate change and the depletion of fossil fuel resources, the development of sustainable and efficient energy technologies has become more critical than ever. Thermoelectric (TE) devices, which can directly convert waste heat into electricity without the need for moving parts or working fluids, have emerged as a promising solution due to their robustness, simplicity, and ability to provide precise thermal management~\cite{TE1,TE2,TE3,TE4,TE5,TE6}. Despite being grounded in the classical Seebeck~\cite{seebeck1} and Peltier effects~\cite{peltier1}, modern TE systems continue to struggle with inherently low conversion efficiency, making the enhancement of the thermoelectric figure of merit ($ZT$) a persistent and central challenge in the field. 

The emergence of nanotechnology has ushered in a transformative era for thermoelectric research, offering powerful strategies to overcome the inherent constraints of conventional bulk materials. At the nanoscale, engineered quantum systems such as quantum dots, carbon nanotubes, molecular junctions, and quantum wires~\cite{c13,c14,c15,c16} have demonstrated remarkable potential in elevating the thermoelectric figure of merit ($ZT$), laying the groundwork for the development of next-generation thermal rectifiers~\cite{rb1} and micro-refrigerators. The incorporation of molecular-scale components into thermoelectric architectures, propelled by recent progress in single-molecule electronics and heterojunction design, has further enriched the field by introducing structurally flexible, electronically tunable, and environmentally sustainable platforms. These innovations have significantly expanded the functional reach of thermoelectric materials, while strengthening the case for low-dimensional~\cite{c10}, scalable energy harvesting solutions. Yet, the field continues to confront critical challenges in realizing stable, high-efficiency performance and achieving fine-tuned control over transport properties.

Optimizing thermoelectric (TE) performance relies on the ability to control four fundamental degrees of freedom: charge, lattice, spin, and orbital. While conventional approaches have primarily targeted phonon suppression and charge carrier optimization to enhance $ZT$, the emergence of spin caloritronics~\cite{spin2,s7,s8,s9,s10,s11,s12,s13,s14,s15,s16} has opened an entirely new frontier in energy conversion. This rapidly evolving field explores the dynamic coupling between heat, charge, and spin, unlocking innovative transport mechanisms and multifunctional device possibilities in solid-state thermal-to-electrical conversion. The introduction of spin degrees of freedom into thermoelectric systems brings transformative potential, offering powerful new channels for tuning transport behavior beyond what charge and lattice dynamics alone can achieve. Spin-driven thermoelectricity typically manifests through three principal pathways. First, spin-related phenomena such as spin entropy, spin fluctuations, and the magnon-drag effect can substantially enhance thermoelectric performance by modifying the conventional charge transport matrix~\cite{Ref18,Ref19,Ref20,Ref21}. Second, the coupling between charge current ($J_C$) and heat current ($J_Q$) becomes magnetically controllable, giving rise to magneto-thermoelectric effects including the magneto-Seebeck effect, its anisotropic counterpart, the Nernst effect, and the anomalous Nernst effect (ANE). Third, and most notably, thermospin effects such as the spin Seebeck effect (SSE)~\cite{Ref22,Ref23,Ref24,Ref25} have redefined the landscape, enabling the conversion of thermally generated spin currents ($J_S$) into measurable charge currents ($J_C$) through the inverse spin Hall effect (ISHE)~\cite{Ref26,Ref27}. Because spin currents inherently dissipate far less energy than charge currents, they present an attractive route toward the development of low-dissipation, high-efficiency thermoelectric devices. In this context, our study offers \textcolor{black}{an exotic} framework that leverages this paradigm by integrating spin-selective quantum interference, topological hopping, and aperiodic potential modulation to engineer next-generation spin-based thermoelectric systems.

Originally formulated to describe the electronic behavior of one-dimensional (1D) polyacetylene chains, the Su–Schrieffer–Heeger (SSH)~\cite{ssh1,ssh2,ssh3,ssh4,ssh5,ssh6,ssh7} model has evolved into a cornerstone of condensed matter theory, celebrated for its elegant tight-binding framework and spontaneous dimerization. Despite its conceptual simplicity, the SSH model has proven to be a fertile ground for uncovering rich physical phenomena such as topological soliton excitations, charge fractionalization, and the manifestation of robust edge states~\cite{Ref2,Ref3,Ref4}. In recent years, extended variants of this model, augmented by periodic modulations in the chemical potential, have been instrumental in probing dynamically driven band topology and quantized topological charge pumping~\cite{Ref12,Ref13,Ref14,Ref15}. Simultaneously, 1D superlattice analogs have garnered significant attention for their capacity to simulate higher-dimensional topological behavior, wherein the modulation phase effectively introduces a synthetic dimension~\cite{Ref16,Ref17}. In particular, the celebrated Haldane model~\cite{Ref20}, which exemplifies the anomalous quantum Hall effect on a two-dimensional (2D) honeycomb lattice, can be effectively mapped onto an extended SSH framework by incorporating cyclically modulated nearest-neighbor (NN) and next-nearest-neighbor (NNN) hopping amplitudes. This configuration facilitates the formation of topologically protected edge states through specific NNN hopping pathways as the modulation phase is tuned. Building upon this rich theoretical landscape, our study pioneers a novel direction by demonstrating, for the first time to the best of our knowledge, that finely tuning the NN and NNN hopping terms within an extended SSH lattice enables high-precision control over spin-resolved transmission spectra for up and down spin channels. This intrinsic tunability not only uncovers the system's latent ability to serve as an efficient quantum spin filter~\cite{spin,sp1,sp2} but also positions it as a transformative platform for exploring spin-selective transport phenomena in low-dimensional topological systems.

In this study, we present a comprehensive investigation into the spin-dependent thermoelectric properties of an extended Su–Schrieffer–Heeger (SSH) model, establishing, to the best of our knowledge, the first demonstration of its potential as an efficient spin thermoelectric generator. A key requirement for achieving high spin thermoelectric efficiency lies in gaining precise control over spin-resolved transmission channels. Our results reveal that a narrow overlap between up and down spin transmission functions significantly enhances spin thermoelectric performance. Additionally, prior studies have highlighted that introducing asymmetry in the transmission line shape can further improve thermoelectric efficiency. Motivated by this, we incorporate a deterministic aperiodic disorder, specifically, Fibonacci-type modulation~\cite{fb1,fb2} , into the onsite energies to induce asymmetry and break transmission symmetry. To provide tunability and capture the rich interplay between competing hopping mechanisms, we include both nearest-neighbor (NN) and next-nearest-neighbor (NNN) hopping amplitudes, each modulated in a cosine form. This dual modulation allows us to finely tune the electronic structure and investigate its impact on both charge and spin thermoelectric transport. Our findings demonstrate significant sensitivity of the thermoelectric response to these parameters, with clear advantages emerging for spin-based performance under optimized conditions. The proposed framework offers a novel route toward designing spin-selective, low-dimensional thermoelectric devices, marking a \textcolor{black}{completely new approach}  step in the intersection of topology, aperiodicity, and spin caloritronics.

The structure of this paper is organized to provide a clear and comprehensive exploration of the proposed framework. Section~I introduces the motivation behind the study, outlines the research objectives, and sets the broader context within the field of spin thermoelectricity. In Section~II, we develop the theoretical foundation, presenting the tight-binding model in detail and employing the Non-Equilibrium Green’s Function (NEGF) formalism~\cite{c19,c20} to evaluate transport and thermoelectric coefficients with high precision. Section~III delivers an in-depth analysis of the results, highlighting how spin-selective transport and Fibonacci pattern intricately affect both charge and spin thermoelectric behavior. Finally, Section~IV encapsulates the major findings, discusses their theoretical and practical implications, and proposes future directions for extending this research. This logical progression ensures a cohesive narrative and reinforces the novelty and applicability of our findings in advancing quantum thermoelectric technologies.

\subsection{Theoretical Framework}

\section{Quantum Design of Thermoelectric Transport: System Architecture and Hamiltonian Framework}

\subsection{Quantum Transport in Ferromagnetic SSH Chains: Tight-Binding Hamiltonian Formalism}

\begin{figure}[ht]
	{\centering \resizebox*{7.3cm}{3.5cm}{\includegraphics{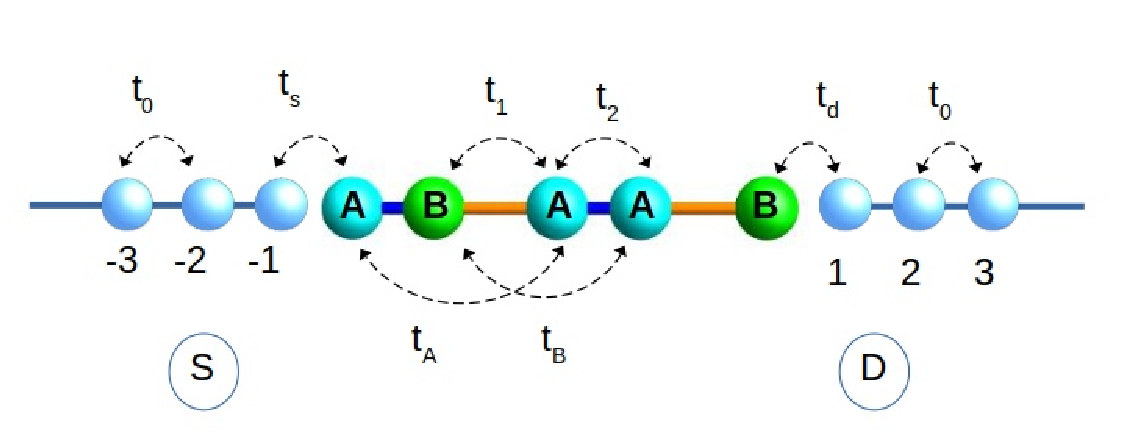}}\par}
			\caption{(Color online). The schematic illustration presents a long range Su-Schrieffer-Heeger (SSH) chain symmetrically connected to source and drain electrodes, where the site energy profile is engineered through a Fibonacci modulation for a specific generation. This unique configuration opens a new frontier in thermoelectric transport by capturing interplay between NN and NNN hopping with aperiodicity.}
			\label{schematic}
		\end{figure}

In this subsection, we outline the thermoelectric configuration based on a Su–Schrieffer–Heeger (SSH) chain that incorporates both nearest- and next-nearest-neighbor hoppings, together with onsite energies arranged according to a Fibonacci aperiodic sequence. The associated tight-binding Hamiltonian captures the essential characteristics of this tailored lattice model. For visualization, Fig.~\ref{schematic} displays a schematic representation of the device: a ferromagnetic (FM) chain containing $N$ magnetic sites, symmetrically connected to two non-magnetic metallic leads, denoted as source (S) and drain (D). This arrangement realizes a spin-dependent thermoelectric junction, where the interplay between aperiodic modulation and engineered hopping terms governs the transport response.

The source and drain electrodes are held at two slightly different temperatures, $T+\Delta T/2$ and $T-\Delta T/2$, thereby generating a finite thermal bias $\Delta T$ across the junction. For analytical transparency and experimental relevance, we confine our discussion to the linear-response regime by assuming $\Delta T$ is sufficiently small. The central ferromagnetic (FM) chain consists of atomic sites each carrying a localized magnetic moment, whose general orientation can be parameterized by the polar angle $\theta_i$ and azimuthal angle $\psi_i$. In this work, we focus on a simplified yet physically pertinent configuration in which all local moments are rigidly aligned along the $+Z$ direction. This alignment stabilizes a uniform ferromagnetic phase characterized by complete spin polarization and a net magnetization pointing along $+Z$. Such an idealized setting, devoid of spin fluctuations or external perturbations, provides a clean platform to examine how an intrinsically spin-polarized background governs the spin-dependent transport and thermoelectric response of the system.

The coupling between mobile electrons and localized magnetic moments arises through standard spin–moment exchange interactions, which generate spin-dependent scattering events that play a decisive role in shaping the transport properties of the system. To faithfully capture the quantum mechanical aspects of this interplay, we employ a tight-binding framework, a minimal yet versatile tool for describing electron dynamics on discrete lattice structures. Within this approach, the total Hamiltonian of the junction is formulated as a sum of distinct components, each representing crucial physical ingredients such as electron hopping, onsite potential variations, and spin-exchange couplings. This unified Hamiltonian serves as the theoretical foundation for probing thermoelectric responses in the setup, allowing us to elucidate how spin-selective scattering channels regulate the concurrent flow of charge and heat in a low-dimensional, magnetically ordered quantum network.

\begin{equation}
H= H_L + H_S + H_D + H_{tn}
\label{eq2}
\end{equation}
where, $H_L$, $H_S$, $H_D$, and $H_{tn}$ represent the sub-Hamiltonians associated with the chain geometry, the source electrode, the drain electrode, and the tunneling connection between the electrodes and the central SSH chain, respectively. For clarity and completeness, we briefly outline the role of each contribution in the following discussion.

The Hamiltonian $H_L$ reads as
\textcolor{black}{
\begin{align}
H_L &= \sum_n \hat{c}_n^\dagger 
       \big(\epsilon_n \otimes \mathbb{I}_1 - \mathbf{h}_n \cdot \boldsymbol{\sigma} \big) \hat{c}_n \nonumber\\
    &\quad + \sum_{n = odd } \Big(
       \hat{c}_n^\dagger \, \hat{t}_1 \, \hat{c}_{n+1} 
       + \mathrm{H.c.} \Big) \nonumber\\
    &\quad + \sum_{n=even} \Big(
       \hat{c}_{n}^\dagger \, \hat{t}_2 \, \hat{c}_{n+1} 
       + \mathrm{H.c.} \Big) \nonumber\\
    &\quad + \sum_{n = odd} \Big(
       \hat{c}_n^\dagger \, \hat{t}_A \, \hat{c}_{n+2} 
       + \mathrm{H.c.} \Big) \nonumber\\
    &\quad + \sum_{n = even} \Big(
       \hat{c}_{n}^\dagger \, \hat{t}_B \, \hat{c}_{n+2} 
       + \mathrm{H.c.} \Big).
\end{align}
}

\textcolor{black}{
where the spinor operator at site $n$ is
\begin{equation}
\hat{c}_n =
\begin{pmatrix}
c_{\uparrow,n} \\
c_{\downarrow,n}
\end{pmatrix},
\end{equation}}

\textcolor{black}{
and the on-site potential $\epsilon_n$ is a $2\times 2$ matrix in spin space:
\begin{equation}
\epsilon_n\otimes \mathbb{I}_1 =
\begin{pmatrix}
\epsilon_n^{\uparrow} & 0 \\
0 & \epsilon_n^{\downarrow}
\end{pmatrix}
\end{equation}
 where 
\begin{equation}
\mathbb{I}_1 =
\begin{pmatrix}
1 & 0 \\
0 & 1
\end{pmatrix}.
\end{equation}}

\textcolor{black}{
The on-site spin-field term $\mathbf{h}_n \cdot \boldsymbol{\sigma}$ in spin space is
\begin{equation}
\mathbf{h}_n \cdot \boldsymbol{\sigma} =
\begin{pmatrix}
h_n \cos \theta_n & h_n \sin \theta_n \, e^{-i \phi_n} \\
h_n \sin \theta_n \, e^{i \phi_n} & -h_n \cos \theta_n
\end{pmatrix}.
\end{equation}}

\textcolor{black}{
The hopping matrices in spin space are proportional to the identity:
\begin{align}
\hat{t}_1 &= t_1 \,\otimes \mathbb{I}_1 =
\begin{pmatrix}
t_1 & 0 \\
0 & t_1
\end{pmatrix}, &
\hat{t}_2 &= t_2 \,\otimes \mathbb{I}_1 =
\begin{pmatrix}
t_2 & 0 \\
0 & t_2
\end{pmatrix}, \nonumber\\
\hat{t}_A &= t_A \,\otimes \mathbb{I}_1 =
\begin{pmatrix}
t_A & 0 \\
0 & t_A
\end{pmatrix}, &
\hat{t}_B &= t_B \,\otimes \mathbb{I}_1 =
\begin{pmatrix}
t_B & 0 \\
0 & t_B
\end{pmatrix}.
\end{align}}

\textcolor{black}{
Here, $c_{\sigma,n}$ ($c_{\sigma,n}^\dagger$) are fermionic annihilation (creation) operators for spin $\sigma$ at site $n$,  all local magnetic moments $h$ are uniformly aligned along the $+Z$ direction, $t_1$ and $t_2$ denote nearest-neighbor hopping amplitudes on alternating bonds, and $t_A$ and $t_B$ denote next-nearest-neighbor hopping amplitudes along alternating sites. The Fibonacci site potential $\epsilon_n$ is arranged according to a deterministic aperiodic sequence generated recursively from initial conditions $F_0 = B$ and $F_1 = A$, with subsequent generations $F_2 = AB$, $F_3 = ABA$, $F_4 = ABAAB$, and so on.
Here, the symbols $A$ and $B$ correspond to two distinct atomic species characterized by onsite potentials $\epsilon_A$ and $\epsilon_B$, respectively.} In order to achieve fine-tuning of the parameters, we consider the form of the NN hopping terms as
\begin{equation}
\begin{split}
    t_1 = t(1+ \delta \cos\beta) \\
    t_2 = t(1-\delta \cos\beta)
\end{split}
\label{eq6}
\end{equation}
and we consider the NNN hoppings in the form 
\begin{equation}
\begin{split}
   t_A = g_a + \eta \cos (\alpha+\beta) \\
   t_B = g_b + \eta \cos (\alpha-\beta)
\end{split}
\label{eq6}
\end{equation}

The sub-Hamiltonians $H_S$ and $H_D$ in Eq.~\eqref{eq2}, corresponding to the source and drain electrodes respectively, are expressed as follows:

\begin{equation}
H_S = \bf \sum_{n\leq -1} a^{\dagger}_{n}\epsilon_{0}a_{n}+\sum_{n\leq -1} \left(a^{\dagger}_{n}t_0a_{n-1}+a^{\dagger}_{n-1}t_0a_{n}\right)
\label{eq6}
\end{equation}
and
\begin{equation}
H_D = \bf \sum_{n\geq 1} b^{\dagger}_{n}\epsilon_{0}b_{n}+\sum_{n\geq 1} \left(b^{\dagger}_{n}t_0b_{n+1}+b^{\dagger}_{n+1}t_0b_{n}\right)
\label{eq7}
\end{equation}
Here, $\mathbf{t}_0$ and $\boldsymbol{\epsilon}_0$ represent the matrices diag$\{t_0, t_0\}$ and diag$\{\epsilon_0, \epsilon_0\}$, respectively. The parameter $t_0$ denotes the nearest-neighbor hopping integral, while $\epsilon_0$ corresponds to the on-site energy within the non-magnetic electrodes.

The tunneling Hamiltonian $H_{tn}$ can be formulated analogously. As illustrated in Fig.~\ref{schematic}, the source electrode is connected to site $1$ of the SSH chain, while the drain is attached to the terminal site $m$. Accordingly, the sub-Hamiltonian $H_{tn}$ is expressed as
\begin{equation}
H_{tn} = \bf \left(a_{-1}^{\dagger}t_s c_1 + c_m^{\dagger}t_db_1 + h.c.\right)
\label{eqtn}
\end{equation}
where ${\bf t_s} = \mathrm{diag}\{t_s, t_s\}$ and ${\bf t_d} = \mathrm{diag}\{t_d, t_d\}$. The parameters $t_s$ and $t_d$ quantify the coupling strengths between the conductor (chain) and the source (S) and drain (D) electrodes, respectively.

\subsection{Theoretical framework}

In this subsection, we present the theoretical framework employed to evaluate the thermoelectric response. The analysis is rooted in the spin-resolved transmission probabilities across the nanojunction. We first formulate the transmission function and then proceed with a step-by-step description of how the corresponding thermoelectric coefficients are extracted from it.

\vskip 0.2cm
\noindent
\emph{\textbf{Spin-Resolved Transmission Characteristics:}} 

To evaluate the spin-dependent transmission probabilities, we adopt the Green’s function formalism, a standard approach that systematically accounts for the effect of the electrodes. Within this framework, the Green’s function of the central chain is expressed as

\begin{equation}
G^r=(G^a)^\dagger =[EI-H_R-\Sigma_S-\Sigma_D]^{-1}
\label{eq8}
\end{equation}

where $\Sigma_S$ and $\Sigma_D$ denote the self-energy matrices associated with the source and drain electrodes, respectively, $I$ represents the identity matrix, and $E$ corresponds to the energy of the incoming electron. These self-energies effectively encode the boundary conditions imposed by the electrodes and capture the escape of electronic states from the central region into the leads.

Using the retarded and advanced Green's functions, $G^r$ and $G^a$, the spin-dependent transmission probabilities are obtained as
\begin{equation}
\tau^{\sigma\sigma^\prime}=Tr[\Gamma_S^\sigma G^r \Gamma_D^{\sigma^\prime} G^a]
\label{eq9}
\end{equation}

here, $\tau^{\sigma\sigma^\prime}$ represents the transmission probability for an incoming electron with spin $\sigma$ to emerge in spin state $\sigma^\prime$. The case $\sigma = \sigma^\prime$ corresponds to spin-conserving transport, while $\sigma \neq \sigma^\prime$ accounts for spin-flip processes. The spin-resolved coupling matrices, $\Gamma_S^\sigma$ and $\Gamma_D^{\sigma^\prime}$, quantifying the hybridization strength between the central region and the source/drain electrodes for the respective spin channels, are evaluated from the self-energy terms according to
\begin{equation}
\Gamma_{S(D)}^{\sigma\sigma^\prime}=i[ \Sigma_{S(D)}^{\sigma\sigma^\prime}- (\Sigma_{S(D)}^{\sigma\sigma^\prime})^\dagger].
\label{eq10}
\end{equation}

By accounting for both spin-preserving and spin-flip contributions, the effective spin-resolved transmission probabilities for the two spin channels can be expressed as:
\begin{equation}
\tau^{\uparrow}=\tau^{\uparrow\uparrow}+ \tau^{\downarrow\uparrow} \hspace{0.5cm} \textup{and} \hspace{0.5cm}
\tau^{\downarrow}=\tau^{\downarrow\downarrow}+ \tau^{\uparrow\downarrow}. 
\label{eq11}
\end{equation}

\vskip 0.2cm
\noindent
\emph{\textbf{Thermoelectric Response Functions:}}
 To unveil the spin-dependent thermoelectric response of our system, we evaluate the Seebeck coefficient, electrical conductance, and electronic thermal conductance by employing the Landauer integrals. These key transport coefficients are determined through the following expressions:
\begin{equation}
S^{\uparrow(\downarrow)}=-\frac{1}{eT}\frac{L^{\uparrow(\downarrow)}_1}{L^{\uparrow(\downarrow)}_0}
\label{eq12}
\end{equation}
\begin{equation}
G^{\uparrow(\downarrow)}= e^2 L^{\uparrow(\downarrow)}_0
\label{eq13}
\end{equation}
\begin{equation}
K^{\uparrow(\downarrow)}_{el}=\frac{1}{T}\left(L^{\uparrow(\downarrow)}_2 -
\frac{(L_{1}^{\uparrow(\downarrow)})^2}{L^{\uparrow(\downarrow)}_0}\right).
\label{eq14}
\end{equation}

\noindent 
The foundational Landauer integrals $L_n$ ($n=0$, $1$, $2$), which form the cornerstone of our evaluation of thermoelectric coefficients, are defined as~\cite{zt1, zt2}
\begin{equation}
L^{\uparrow(\downarrow)}_{n}=-\frac{1}{h} \int \tau^{\uparrow(\downarrow)}
\left(\frac{\partial f (E)}{\partial E}\right) \left(E-E_F \right)^{n}\, {d}E
\label{eq15}
\end{equation}

in these relations, $f(E)$ represents the Fermi–Dirac distribution function, while $E_F$ corresponds to the equilibrium Fermi energy. Building upon the components defined in Eqs.\ref{eq13}–\ref{eq15}, one can systematically derive the thermoelectric quantities for both charge and spin sectors. Specifically, the Seebeck coefficients $(S_C, S_S)$, electrical conductances $(G_C, G_S)$, and the electronic contribution to the thermal conductance $(K)$ are obtained as follows\cite{zt1, zt2}:
\begin{equation}
	S_{C}=\frac{S^{\uparrow}+S^{\downarrow}}{2}\hspace{0.5cm} \textup{and} \hspace{0.5cm}
	S_{S}=S^{\uparrow}-S^{\downarrow}
	\label{eq16}
\end{equation}
\begin{equation}
	G_{C}=G^{\uparrow}+G^{\downarrow}\hspace{0.5cm} \textup{and} \hspace{0.5cm}   
	G_{S}=G^{\uparrow}-G^{\downarrow}
	\label{eq17}
\end{equation}
\begin{equation}
	K_{C}=K_{S}=K^{\uparrow}+K^{\downarrow}=K.
	\label{eq18}
\end{equation}

For a physically motivated simplification, we emphasize that the considered chain is of nanoscale size. In such a regime, the phononic contribution to thermal conductance, $K_{ph}$, becomes negligibly small. Consequently, its effect can be disregarded, enabling us to treat the electronic thermal conductances as identical, i.e., $K_e = K_C = K_S = K$, without loss of generality.

Finally, the charge and spin thermoelectric figures of merit, represented as $Z_C T$ and $Z_S T$, are obtained from the standard relations:
\begin{equation}
Z_CT=\frac{S_C^2 G_C T} {K} \hspace{0.5cm} \textup{and} \hspace{0.5cm} Z_ST=\frac{S_S^2 G_S T} {K}.
\label{eq19}
\end{equation}

\section{Numerical Results and Discussion}
\label{sec:results}

In this section, we present a comprehensive numerical investigation of the thermoelectric transport properties of the proposed model, revealing unexplored parameter regimes that yield significantly enhanced thermoelectric efficiency, as reflected in large values of the figure of merit ($ZT$). Our analysis highlights how the subtle interplay among hopping strengths, Fibonacci modulations of the onsite potential, and tailored structural arrangements can be strategically tuned to optimize both electronic transport and thermoelectric response, thereby opening new avenues for the design of efficient low-dimensional thermoelectric devices. Except for the cases where system size variation is explicitly considered, the chain length is fixed at $N=8$, arranged according to the specific sequence $ABAABABA$. The specific values and corresponding units of the remaining parameters are presented within the discussion sections of the subsequent subsections.

\subsection{Spectral Profile of the Transmission Function}

\begin{figure}[ht]
{\centering \resizebox*{7.5cm}{7.5cm}{\includegraphics{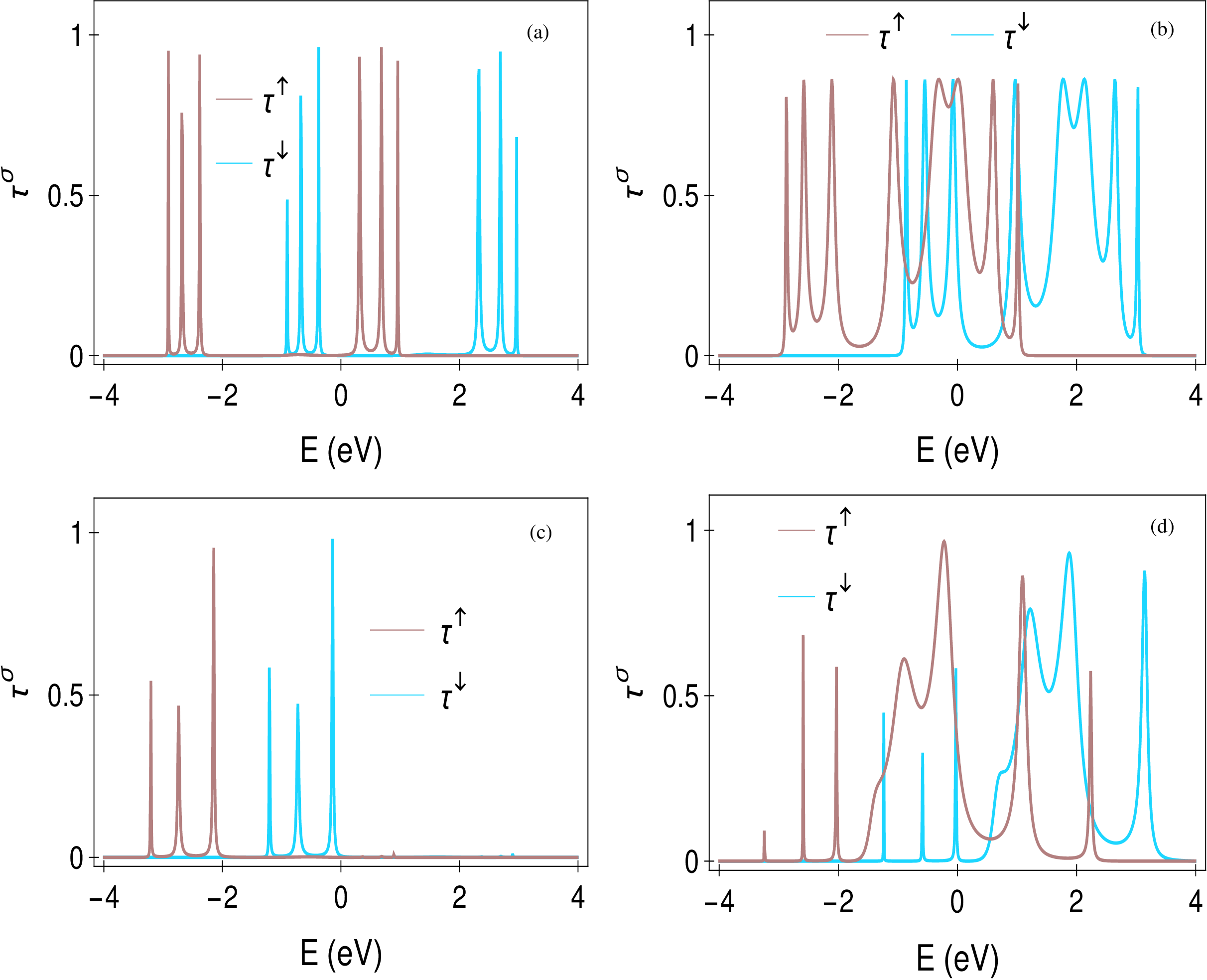}}\par}
\caption{(Color online). The plot illustrates the transmission function as a function of energy under various hopping configurations. Panel (a) displays fully spin-resolved transmission spectra, highlighting complete separation between up and down spin channels under nearest-neighbor (NN) hopping. In contrast, panel (b) shows partial overlap between the spin-resolved channels, still within the NN hopping-only framework. Panel (c) incorporates next-nearest-neighbor (NNN) hopping and again reveals complete spin-channel separation, indicating the strong influence of longer-range hopping. Finally, panel (d) includes both NN and NNN hoppings, resulting in overlapping transmission lines, thereby capturing the interplay between the two hopping mechanisms and their collective impact on spin-resolved transport.}
\label{fig1}
\end{figure}

We begin our analysis by examining the behavior of the energy-dependent transmission function $\mathcal{T}(E)$, tuning the NN and NNN hopping parameters in a precise manner. We find that when we only consider the inter-cell and intra-cell NN hoppings $t_1$ and $t_2$ for the SSH chain, we can separate the up-spin and down-spin channels. In order to achieve fine-tuning of the parameters, we consider the form $t_1 = t(1+ \delta \cos\beta)$ and $t_2 = t(1-\delta \cos\beta)$. As shown in Fig.~\ref{fig1}(a), for $\delta = 0.8$ and $\beta = \frac{4\pi}{5}$, we obtain complete spin separation. However, when $\delta = 0.5$ and $\beta = \frac{\pi}{2}$, the spin channels overlap significantly [see Fig.~\ref{fig1}(b)]. To gain more refined control over the transmission function, we consider the NNN hoppings in the form $t_M = g_a + \eta \cos (\alpha+\beta)$ and $t_N = g_b + \eta \cos (\alpha-\beta)$, where $t_M$ and $t_N$ represent hopping between odd and even sites, respectively. The inclusion of four parameters allows better tuning of $t_M$ and $t_N$, offering more precise control over the spin transmission channels. In Fig.~\ref{fig1}(c), we set $\delta = 0.8$, $\beta = 0.7\pi$, $g_a = g_b = 0.1$, $\alpha = \frac{3\pi}{4}$, and $\eta = 0.5$, which yields well-separated spin channels. On the other hand, Fig.~\ref{fig1}(d) displays overlapping transmission spectra for up and down spins using $\delta = 0.8$, $\beta = 0.7\pi$, $g_a = g_b = 0.5$, $\alpha = \frac{\pi}{2}$, and $\eta = 0.5$. We adopt a Fibonacci-type modulation for the onsite energies, where consecutive site energies are taken as $-0.5$ and $0.5$~eV, respectively. The system is constructed with order number $5$, consisting of $8$ sites. Higher-order results are also discussed later. From this transmission profile, we understand that tuning the NN and NNN hopping terms allows us to manipulate spin transmission efficiently. Using cosine-modulated hopping amplitudes with adjustable parameters provides a fine control mechanism for spin channel separation. Since high spin thermoelectric efficiency requires minimal overlap between up and down spin transmissions, this model shows strong potential for designing efficient spin thermoelectric devices.
 
\begin{figure}[ht]
{\centering \resizebox*{7.5cm}{4cm}{\includegraphics{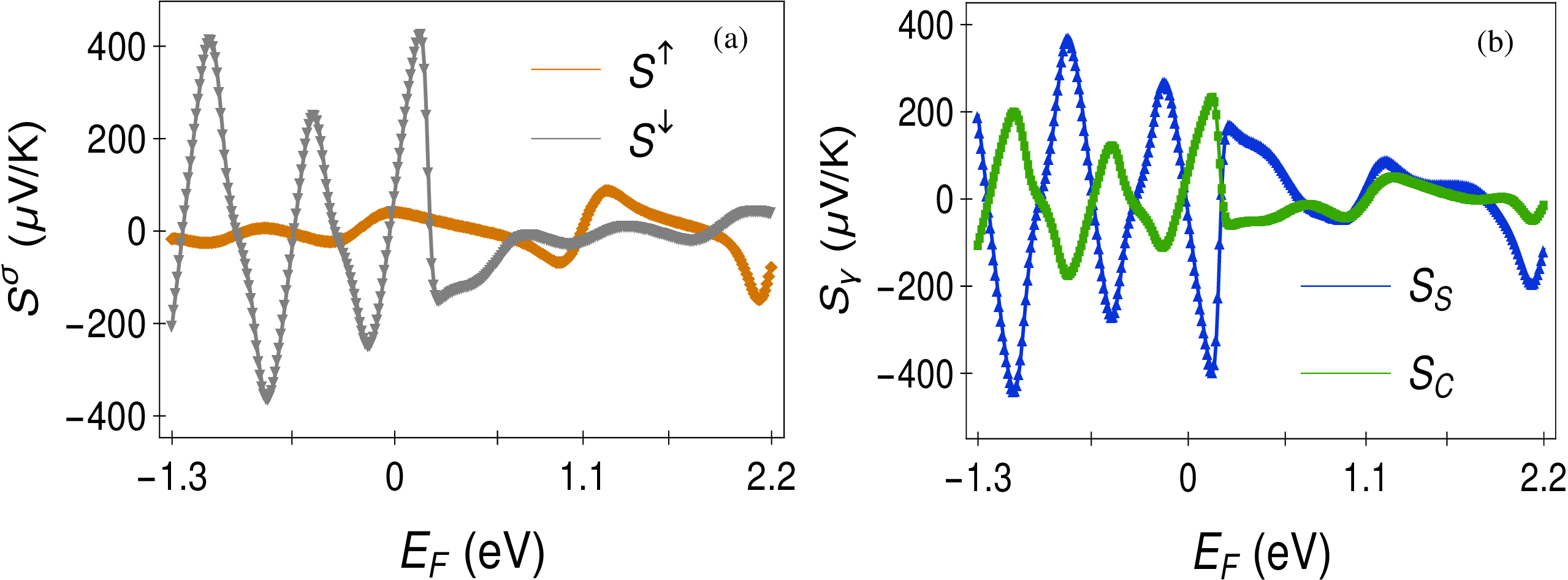}}\par}
\caption{(Color online). Seebeck coefficient as a function of Fermi energy is shown. Panel (a) highlights the spin-resolved Seebeck response, with distinct contributions from up-spin and down-spin channels. Panel (b) presents the corresponding charge and spin Seebeck coefficients, capturing the overall thermoelectric behavior across both transport modes.}
\label{fig2}
\end{figure}

\subsection{Seebeck Coefficient}
We next calculate the Seebeck coefficients for both up and down spin channels in the presence of both nearest-neighbor (NN) and next-nearest-neighbor (NNN) hoppings, followed by the evaluation of the charge and spin Seebeck coefficients using Eq.~\ref{eq16}. To elucidate the origin and behavior of both charge and spin Seebeck responses, along with their spin-resolved components, we analyze the central part of the Landauer integral $L_1$, expressed as $\tau^\sigma(E)\left(\frac{\partial f}{\partial E}\right)(E-E_F)$ for each spin channel. The presence of the $(E-E_F)$ term naturally allows $S^{\sigma}$ to take on both positive and negative values depending on the location of the Fermi energy $E_F$. As shown in Fig.~\ref{fig2}(a), the spin Seebeck coefficients display notably asymmetric profiles in both magnitude and sign, with the up and down spin components contributing oppositely over a wide energy range. These features are encapsulated in the total charge and spin Seebeck coefficients, $S_{\gamma}$ (with $\gamma = C,S$), presented in Fig.~\ref{fig2}(b). Due to the negative sign in Eq.~\ref{eq12}, the Seebeck coefficient exhibits an inverse relation with the spin-resolved transmission function $\tau^\sigma$, such that rising edges in $\tau^\sigma$ correspond to trailing features in $S^\sigma$, and vice versa. This interplay leads to well-separated peaks in the Seebeck coefficients, resulting from the offset in the rising and trailing edges of $\tau^{\uparrow}$ and $\tau^{\downarrow}$. Interestingly, Fig.~\ref{fig1}(d) highlights that within the energy range $-1.3$ to $0$~eV, the points where $\tau^{\uparrow}$ intersects $\tau^{\downarrow}$ coincide with sharp increases in $S^{\uparrow}$, as also seen in Fig.~\ref{fig2}(a). Owing to their largely opposite characteristics for a given $E_F$, a significant spin Seebeck coefficient naturally emerges, clearly surpassing the charge counterpart, as illustrated in Fig.~\ref{fig2}(b). This behavior can be directly traced to the profiles of $S^{\uparrow}$ and $S^{\downarrow}$, which dictate the contrasting nature of $S_S$ and $S_C$. The charge Seebeck coefficient reaches a maximum of about $200\,\mu$V/K, whereas the spin counterpart attains nearly $400\,\mu$V/K. A detailed examination across the full Fermi energy window confirms that $S_S$ consistently dominates over $S_C$. For this analysis, we have adopted the same hopping parameters as in Fig.~\ref{fig1}(d), which we retain throughout the rest of the study unless otherwise noted. As the Seebeck coefficient enters quadratically in the numerator of $Z_{\gamma}T$ (Eq.~\ref{eq19}), its magnitude plays a critical role in determining the thermoelectric efficiency. The observed large $S_S$ and comparatively small $S_C$ thus have a direct and pronounced impact on $Z_ST$ and $Z_CT$, as will be discussed in the subsequent subsection. These findings emphasize how a controlled imbalance between the spin-resolved transmission functions, along with their minimal overlapping window, can be strategically harnessed to realize a pronounced spin Seebeck response.

\subsection{Electrical and thermal conductances} 

\begin{figure}[ht]
{\centering \resizebox*{7.5cm}{4cm}{\includegraphics{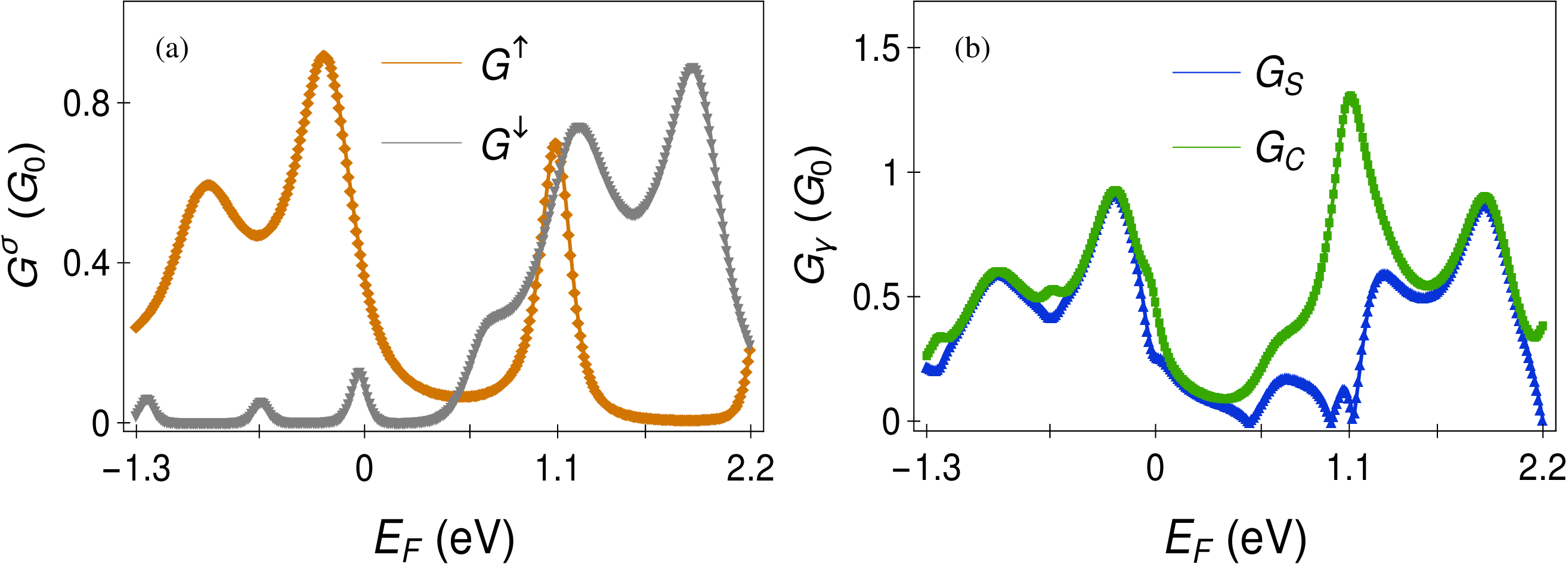}}\par}
\caption{(Color online). Spin-resolved electronic conductance as a function of Fermi energy is presented in panel (a), highlighting distinct contributions from up- and down-spin carriers. Panel (b) contrasts the total charge and spin conductances, providing insight into the system’s thermoelectric response under varying transport channels.}
\label{fig3}
\end{figure}

In this sub-section, we explore the behavior of two additional thermoelectric quantities, electrical and thermal conductances, that are fundamentally linked to the efficiency of energy conversion in nanoscale systems. To maintain consistency and enhance interpretability, we focus on the same energy window as before, spanning from $-1.3\,$eV to $2.2\,$eV. By nature, conductance is inherently a positive-definite quantity, and our analysis confirms this expectation for both spin-resolved electrical conductances. As depicted in Fig.~\ref{fig3}(a), the calculated spin conductances, $G^\uparrow$ and $G^\downarrow$, remain positive throughout the entire energy range, indicating robust and spin-dependent transport channels under the combined influence of nearest-neighbor and next-nearest-neighbor hopping mechanisms.
 
The conductance profile exhibits a strong correlation with the transmission function, resulting in multiple energy regions where the spin-resolved conductances $G^{\uparrow}$ and $G^{\downarrow}$ overlap significantly. Given that conductance remains strictly positive by definition, the application of Eq.~\ref{eq17} yields insightful comparisons between the charge and spin components. Specifically, it is observed that the charge conductance $G_C$ consistently surpasses its spin counterpart $G_S$, a trend quantitatively evident in Fig.~\ref{fig3}(b) where $G_C \approx 1.5~G_0$ and $G_S \approx 0.8~G_0$ \textcolor{black}{($G_0 = 2 e^2/h$)}. This elevated value of $G_C$ contributes favorably to the charge figure of merit $Z_CT$. Nonetheless, owing to the significantly higher spin Seebeck coefficient compared to the charge Seebeck coefficient, the thermoelectric efficiency in the spin channel is found to be superior, with the enhanced $S_S$ playing a pivotal role in boosting the spin-dependent thermoelectric performance beyond that of the charge-based counterpart.

\begin{figure}[ht]
{\centering \resizebox*{7.5cm}{5cm}{\includegraphics{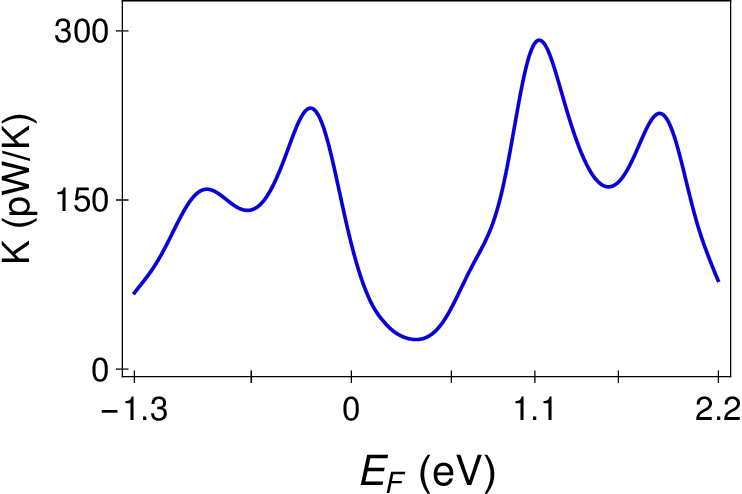}}\par}
\caption{(Color online). Thermal conductance as a function of Fermi energy, the parameters used in this analysis are consistent with those in Fig.~\ref{fig1} (d). This offers a complete perspective on the thermoelectric transport characteristics of the system.}
\label{fig4}
\end{figure}
  
The behavior of thermal conductance is illustrated in Fig.~\ref{fig4}, where it is evident that both charge and spin channels yield identical values of $K$, implying that thermal conductance does not independently influence $Z_{C}T$ and $Z_{S}T$. Nonetheless, minimizing $K$ remains critical for enhancing the figure of merit, as it appears in the denominator of Eq.~\ref{eq19}. In our numerical analysis, the peak value of $K$ is found to be approximately $290\,$pW/K. The variation of $K$ as a function of Fermi energy bears strong resemblance to the trend of charge electrical conductance $G_C$, yet their exact proportionality does not hold across the entire energy range. This deviation from proportionality indicates the \textcolor{black}{Wiedemann-Franz (WF) law~\cite{wf1,wf2}is not getting exactly followed here}, a feature often considered favorable for achieving higher thermoelectric efficiency in nanoscale systems and also supported by numerous recent studies. It is important to note that the current study focuses exclusively on the electronic contribution to thermal conductance. \textcolor{black}{We performed an order-of-magnitude estimate of the phonon thermal conductance for an SSH chain and obtained a value of approximately $29$ pW/K. Though electronic thermal conductance is around $290$ pW/K, an order of magnitude larger. In addition, our survey of relevant literature indicates that reported values of phonon thermal conductance in similar nanoscale chains  usually resides within the range of $5$–$50$ pW/K ~\cite{kph1}. Thus we  omit the phononic component as we aim towards qualitative enhancement of TE \textcolor{black}{coefficient}.}

\subsection{Charge and spin figure of merits ($Z_CT$, $Z_ST$)} 

\begin{figure}[ht]
{\centering \resizebox*{7.5cm}{5cm}{\includegraphics{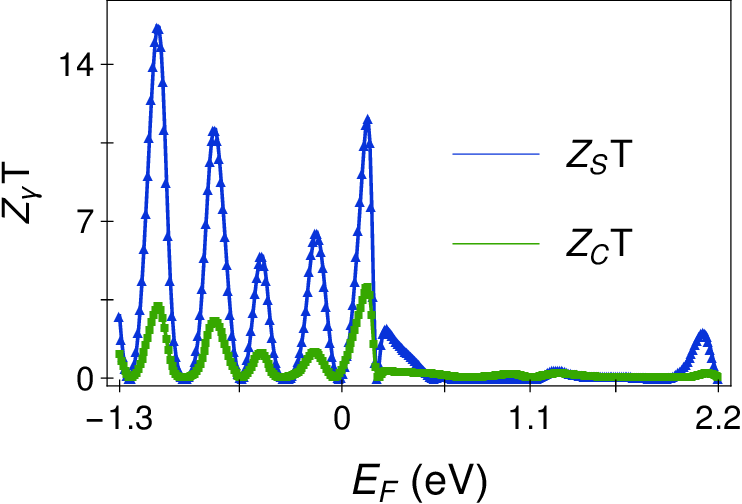}}\par}
\caption{(Color online). Thermoelectric figure of merit ($ZT$) as a function of Fermi energy. Both spin and charge $ZT$ are shown and the parameters used are consistent with those in Fig.~\ref{fig1} (d).
}
\label{fig5}
\end{figure}

Building upon the comprehensive analysis of all the relevant thermoelectric quantities discussed earlier, we now turn our attention to the behavior of the charge and spin figures of merit, as depicted in Fig.~\ref{fig5}, where the blue and green curves correspond to spin and charge responses, respectively. A distinct and consistent enhancement of $Z_{S}T$ over $Z_{C}T$ is observed throughout the Fermi energy window ranging from $-1.3$ to $2.2\,$~eV. Quantitatively, \textcolor{black}{our study offers a qualitative estimate of the thermoelectric parameters,} while $Z_{C}T$ exhibits a peak around $4$, the spin counterpart, $Z_{S}T$, attains a significantly higher value close to $17$. This substantial improvement in $Z_{S}T$ is directly linked to the elevated spin Seebeck coefficient $S_S$ in comparison to its charge analog $S_C$, a behavior which stems from the inherent asymmetry between the up and down spin transmission functions. As previously highlighted, the energy regions near the crossing points of these spin-resolved transmission spectra are particularly conducive to achieving such pronounced enhancements in spin thermoelectric performance. With this understanding, we now delve deeper into how nearest-neighbor hopping (NN) and next-nearest-neighbor hopping (NNN) integrals intricately modulate the energy conversion efficiency, providing deeper insights into the tunability and design principles of such spin-resolved thermoelectric systems.

\subsubsection*{Impact of Nearest-Neighbor Hopping}

\subsection{Tuning Maximum Thermoelectric Quantities via $\delta$}

In this sub-section, we explore the influence of $\delta$, the modulation strength of the cosine term in the nearest-neighbor hopping, on the spin and charge thermoelectric parameters. To this end, we compute the maximum values of these thermoelectric quantities by scanning over a sufficiently broad range of Fermi energies and analyze their variation with $\delta$, \textcolor{black}{we find in our study that $\delta$ primarily controls the line shape of the transmission function}. The primary motivation behind this analysis is to assess whether the model can yield consistently high $Z_\gamma T$ values across a wide parameter space or if it performs optimally only for specific fine-tuned values. This distinction is crucial in determining the robustness of the model as a potential thermoelectric generator. Our findings reveal that the maximum values of $Z_\gamma T$ remain appreciably high (e.g., $\ge 3$) throughout the entire considered range of $\delta$, confirming its stability and efficiency. Notably, the spin figure of merit, $Z_S T^{\text{max}}$, reaches a peak of approximately $\sim 18$, while its charge counterpart, $Z_C T^{\text{max}}$, attains around $\sim 4$. To provide a comprehensive perspective, we further examine the maximum Seebeck coefficients, where $S_S^{\text{max}}$ achieves values close to $\sim 470\,\mu$V/K and $S_C^{\text{max}}$ around $\sim 270\,\mu$V/K, again highlighting the dominance of the spin channel. We also analyze the maximum electronic conductances, finding that $G_C^{\text{max}} \approx 1.4\,G_0$ surpasses $G_S^{\text{max}} \approx 0.6\,G_0$. Finally, the thermal conductance exhibits a peak value of $K_{\text{max}} \approx 321$\,pW/K. The parameter values employed for this analysis are $\beta = 0.3\pi$, $g_a = 0.3$, $g_b = 0.3$, $\eta = 0.5$, and $\alpha = 0.2\pi$. However, it is important to note that such elevated figure-of-merit values are not exclusive to this parameter set; similar results are observed for other configurations as well. Overall, the robustness of $Z_\gamma T$ with respect to variations in $\delta$ underscores the potential of this model as a promising and efficient thermoelectric generator.
 
\begin{figure}[ht]
{\centering \resizebox*{8cm}{8cm}{\includegraphics{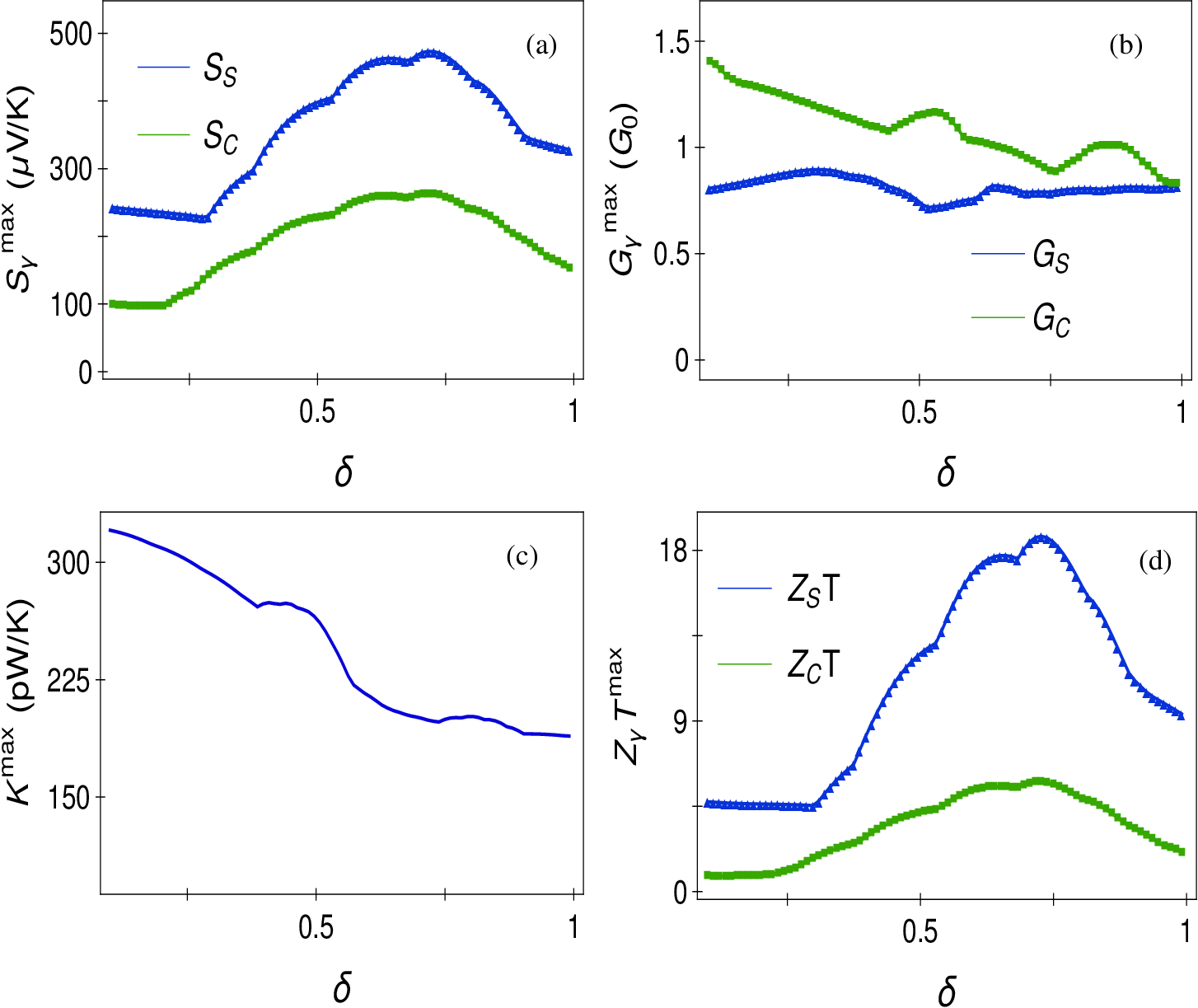}}\par}
\caption{(Color online). The plot presents the maximum values of various thermoelectric quantities as function of the $\delta$. Panel (a) shows the variation of the maximum spin and charge Seebeck coefficients, while panel (b) displays the corresponding maximum conductance values. Panel (c) illustrates the behavior of the maximum thermal conductance for both spin and charge channels. Finally, panel (d) depicts the maximum thermoelectric figure of merit ($ZT$) for spin and charge transport as a function of $\delta$, offering a comprehensive overview of how each component responds to parameter variation.}
\label{fig6}
\end{figure}

\subsection{Dependence of Maximum Thermoelectric Parameters on $\beta$} 

\begin{figure}[ht]
{\centering \resizebox*{8cm}{8cm}{\includegraphics{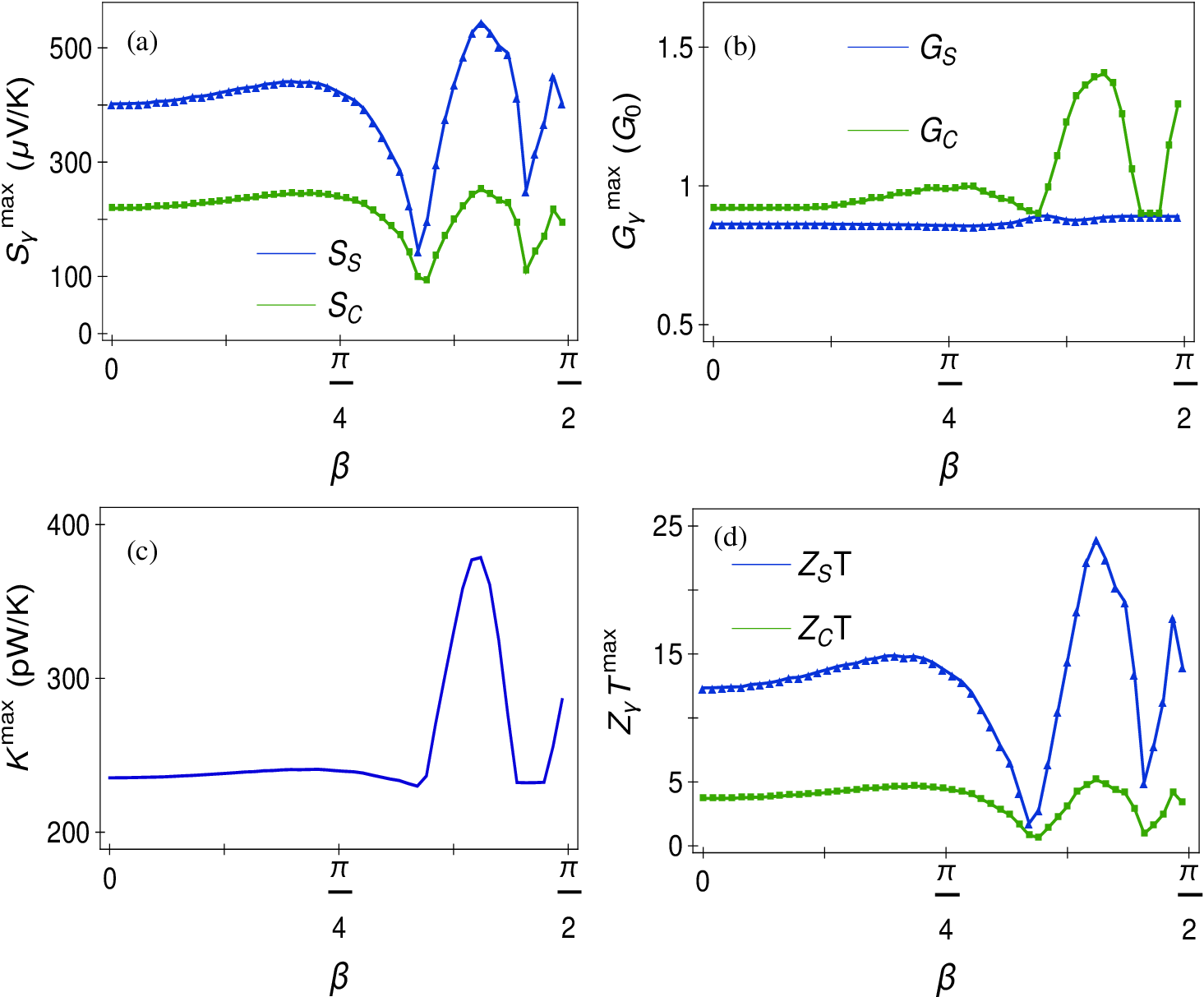}}\par}
\caption{(Color online). This plot illustrates the maximum values of thermoelectric quantities as function of the $\beta$. Panel (a) shows the variation of the maximum spin and charge Seebeck coefficients, while panel (b) presents the corresponding conductance values. Panel (c) depicts the variation in thermal conductance, and panel (d) shows the maximum thermoelectric figure of merit ($ZT$) for both spin and charge channels. All panels highlight the dependence on $\beta$, offering a comprehensive view of how spin and charge components evolve with this parameter.
}
\label{fig7}
\end{figure}

In this sub-section, we investigate the dependence of spin and charge thermoelectric parameters on $\beta$, the phase angle associated with the cosine modulation in the nearest-neighbor hopping. While $\delta$ primarily governs the shape of the transmission function, $\beta$ plays a crucial role in determining the relative alignment of the up and down spin transmission spectra. As such, $\beta$ directly influences the degree of separation or overlap between spin channels, and therefore emerges as a significant tuning parameter for spin filtration. This sensitivity to channel separation implies that fine control of $\beta$ may lead to enhanced spin thermoelectric efficiency. To quantify this effect, we compute the maximum values of thermoelectric parameters over a sufficiently wide range of Fermi energies and analyze their variation with $\beta$. The primary objective here is to examine how the extent of overlap between spin-resolved transmission functions impacts $Z_\gamma T$. We perform this analysis over the interval $0 \le \beta \le \pi/2$, beyond which the spin channels become completely separated with no overlap. Our results show that $Z_\gamma T$ remains high across the entire range, typically exceeding 3. Specifically, we find that the maximum spin figure of merit, $Z_S T^{\text{max}} $, reaches up to $\sim 21$, while the corresponding charge figure of merit, $Z_C T^{\text{max}} $, attains a value around $\sim 4$. To complete the picture, we also examine the variations of other thermoelectric quantities with $\beta$. The maximum spin Seebeck coefficient, $S_S^{\text{max}}$, reaches $\sim 511\,\mu$V/K, while its charge counterpart, $S_C^{\text{max}}$, attains $\sim 270\,\mu$V/K, once again indicating the superiority of the spin response. In the case of conductance, $G_C^{\text{max}} \approx 1.5\,G_0$ and $G_S^{\text{max}} \approx 1.0\,G_0$. The maximum thermal conductance, $K^{\text{max}}$, is found to be approximately $358$\,pW/K. The parameter values used in this study are $\delta = 0.7$, $g_a = 0.3$, $g_b = 0.3$, $\eta = 0.5$, and $\alpha = 0$. Notably, such high values of the figure of merit are not unique to this specific parameter set, as similar trends are observed for other combinations as well. This investigation demonstrates that $\beta$ is a powerful \textcolor{black}{controlling} parameter for optimizing thermoelectric efficiency, and highlights the critical role of spin channel overlap in enhancing $Z_\gamma T$.

\subsection{Variation of maximum values of thermoelectric parameters with $g_a$} 

\begin{figure}[ht]
{\centering \resizebox*{8cm}{8cm}{\includegraphics{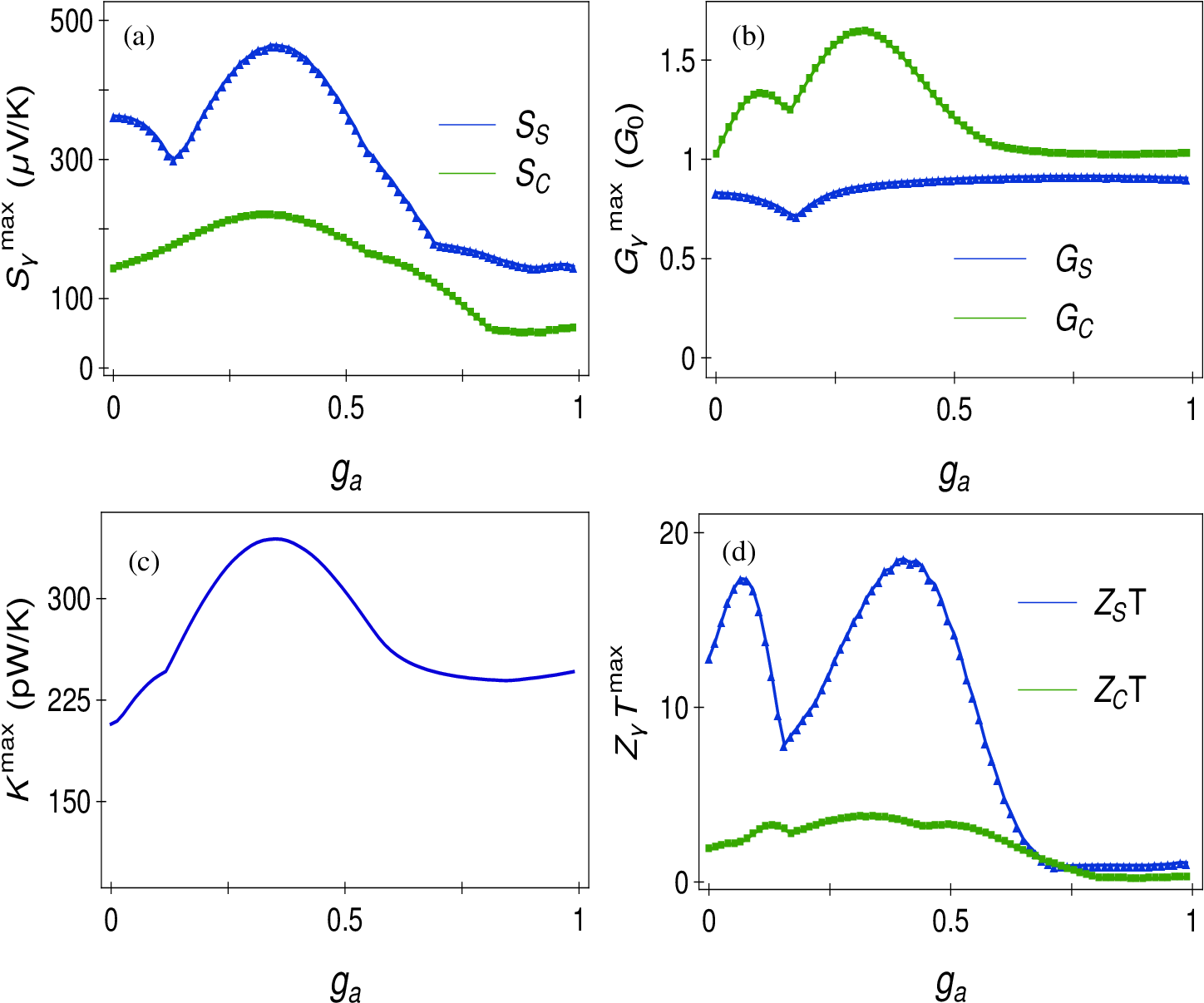}}\par}
\caption{(Color online). This plot presents the maximum values of thermoelectric quantities as function of the $g_a$. Panel (a) displays the variation in the maximum spin and charge Seebeck coefficients, while panel (b) shows the corresponding maximum conductance values. Panel (c) depicts the variation in maximum thermal conductance, and panel (d) illustrates the maximum thermoelectric figure of merit ($ZT$) for both spin and charge channels. The results highlight how each component evolves with $g_a$, providing a comprehensive understanding of the system's thermoelectric response.}
\label{fig8}
\end{figure}

In a new stride towards advancing spin caloritronics, we uncover a novel tuning mechanism for thermoelectric optimization by exploring the influence of $g_a$, the constant term in next-nearest-neighbor (NNN) inter-cell hopping. The NNN hopping is carefully engineered as $t_M = g_a + \eta \cos (\alpha+\beta)$ and $t_N = g_b + \eta \cos (\alpha-\beta)$, where $t_M$ and $t_N$ denote hoppings between odd and even sites, respectively. By decomposing the hopping into four tunable parameters—$g_a$, $g_b$, $\eta$, and $\alpha$, our approach offers unprecedented control over the transport landscape, marking a significant departure from conventional models. Strikingly, $g_a$ emerges as a key modulator of spin-dependent transmission spectra, inducing far more pronounced asymmetries in line shapes compared to what was achievable with $\delta$. This enhanced asymmetry, as our results vividly demonstrate, directly feeds into the amplification of the Seebeck coefficient and, consequently, the figure of merit $Z_\gamma T$. To rigorously assess this impact, we compute the peak values of thermoelectric quantities by scanning a broad Fermi energy range as a function of $g_a$. Our findings are both consistent and compelling: across the entire parameter space of $g_a$, $Z_\gamma T$ values remain robustly high, frequently exceeding 3, with $Z_S T^{\text{max}}$ reaching an impressive $\sim 18$ and $Z_C T^{\text{max}}$ around $\sim 3$. A detailed breakdown of associated thermoelectric parameters further reinforces this trend, $S_S^{\text{max}}$ peaks near $\sim 470\,\mu$V/K, while $S_C^{\text{max}}$ follows at approximately $\sim 240\,\mu$V/K, highlighting the dominance of spin Seebeck response. In the realm of electrical conductance, we observe $G_C^{\text{max}} \approx 1.6$~$G_0$ outpacing $G_S^{\text{max}} \approx 1$~$G_0$, while the maximum thermal conductance stabilizes at $K^{\text{max}} \approx 343$\,pW/K. These results are obtained for a representative parameter set: $\delta = 0.7$, $\beta = 0$, $g_b = 0.3$, $\eta = 0.5$, and $\alpha = 0$, although we emphasize that the emergence of large $Z_\gamma T$ values is not unique to this choice similar performance is replicated across diverse configurations. Taken together, our study not only introduces a powerful new axis of control via $g_a$, but also sets a new benchmark for thermoelectric efficiency in aperiodic systems by leveraging spin-resolved transmission asymmetry, a paradigm-shifting advance in the design of high-performance thermoelectric generators.

\subsection{Dependence of Maximum Values of Thermoelectric Parameters on $g_b$}
 
\begin{figure}[ht]
{\centering \resizebox*{8cm}{8cm}{\includegraphics{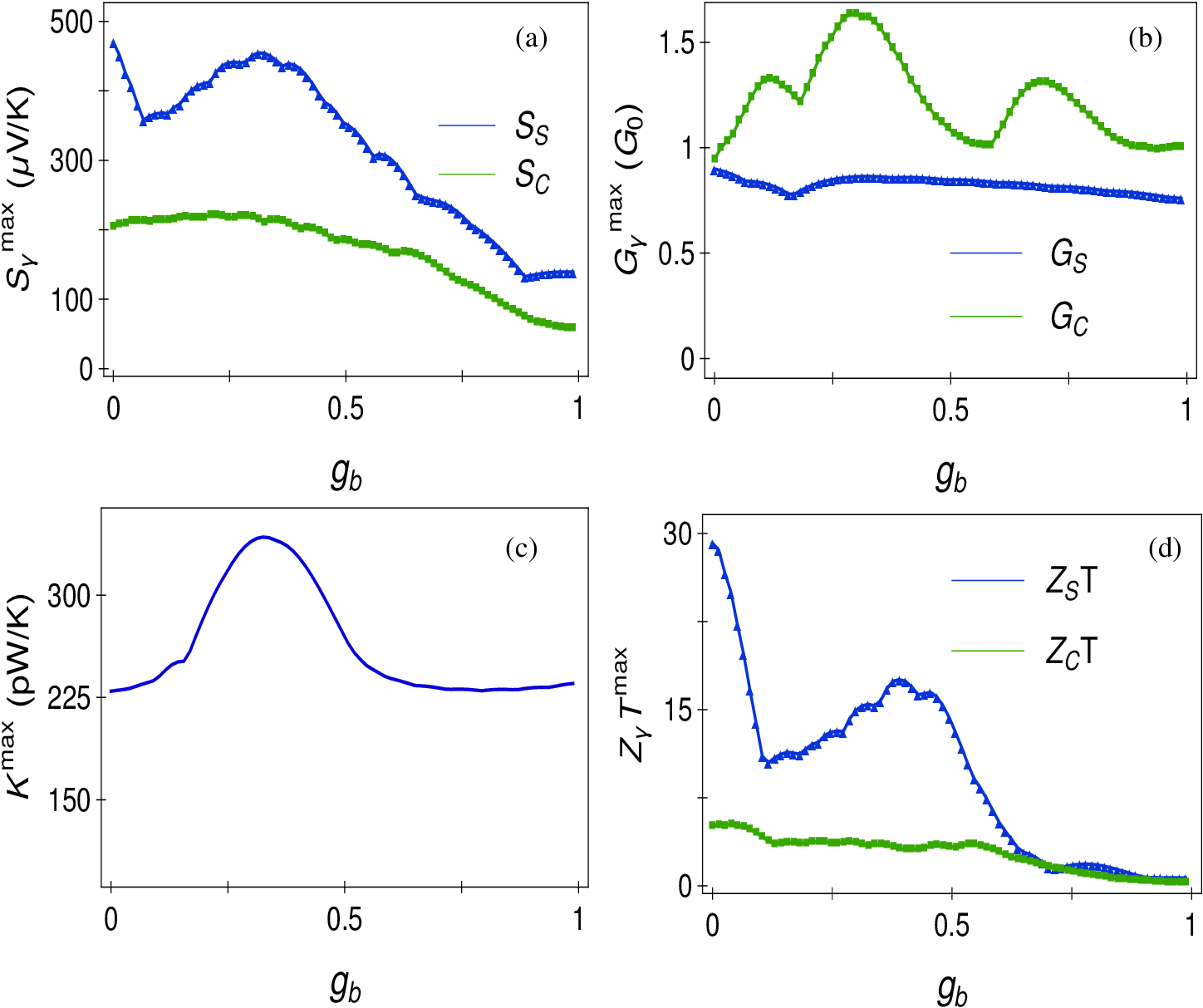}}\par}
\caption{(Color online). This plot illustrates the maximum values of thermoelectric quantities as functions of $g_b$. Panel (a) shows the variation of the maximum spin and charge Seebeck coefficients, while panel (b) presents the corresponding conductance values. Panel (c) depicts the maximum thermal conductance for both spin and charge channels, and panel (d) displays the variation of the maximum thermoelectric figure of merit ($ZT$). Together, these panels provide comprehensive insights into how the spin and charge components evolve with $g_b$, highlighting the tunability of the system's thermoelectric performance.}
\label{fig9}
\end{figure}

Pushing the boundaries of thermoelectric design, we delve into the transformative role of $g_b$, the constant term in next-nearest-neighbor intra-cell hopping, and its profound impact on spin and charge thermoelectric performance. The intra-cell NNN hopping is introduced through the relation $t_N = g_b + \eta \cos (\alpha - \beta)$, representing the coupling between even sites. By structuring the hopping in terms of three distinct tunable parameters $g_b$, $\eta$, and $\alpha$, our framework allows precise modulation of transport asymmetry, a critical ingredient for high-efficiency thermoelectric response. Mirroring our analysis of $g_a$, we systematically scan the Fermi energy and map the peak thermoelectric quantities as functions of $g_b$, \textcolor{black}{this impacts on both the lineshape and the position of up an down spin transmission channels}. While one might initially expect $g_a$ and $g_b$ to exert similar influence, given the analogous mathematical forms of $t_M$ and $t_N$, our results reveal a striking distinction. The figure of merit $Z_\gamma T$ exhibits significantly stronger enhancement with $g_b$ than with $g_a$, as evident in Fig.~\ref{fig9} compared to Fig.~\ref{fig8}, thus underscoring the pivotal role of intra-cell NNN hopping processes, an effect uniquely emergent within the SSH-NNN  framework. Across the full range of $g_b$, we consistently observe high values of $Z_\gamma T$ (e.g., $\ge 3$), with $Z_S T^{\text{max}}$ peaking near $\sim 29$ and $Z_C T^{\text{max}}$ approaching $\sim 4$. To deepen our understanding, we also evaluate associated thermoelectric parameters. The spin Seebeck coefficient $S_S^{\text{max}}$ reaches an impressive $\sim 470$~$\mu V/K$, again leading over the charge counterpart $S_C^{\text{max}} \sim 240$~$\mu V/K$, highlighting the dominant spin contribution. Electrical conductance exhibits a complementary trend, with $G_C^{\text{max}} \approx 1.6$~$G_0$ exceeding $G_S^{\text{max}} \approx 1$~$G_0$, and thermal conductance saturating at $K^{\text{max}} \approx 343$~pW/K. These results are obtained for a representative set of parameters: $\delta = 0.7$, $\beta = 0$, $g_a = 0.3$, $\eta = 0.5$, and $\alpha = 0$, though our broader explorations confirm that high-performance thermoelectric behavior is not confined to this specific regime. Ultimately, our findings establish $g_b$ as a more effective tuning knob than $g_a$, and reveal that intra-cell NNNH hopping is a potent driver of enhanced $Z_\gamma T$, a key insight that opens new avenues in the strategic engineering of next-generation thermoelectric materials.

\subsection{Exploring the Maximum values of Thermoelectric Parameters with $\eta$}

\begin{figure}[ht]
{\centering \resizebox*{8.0cm}{8cm}{\includegraphics{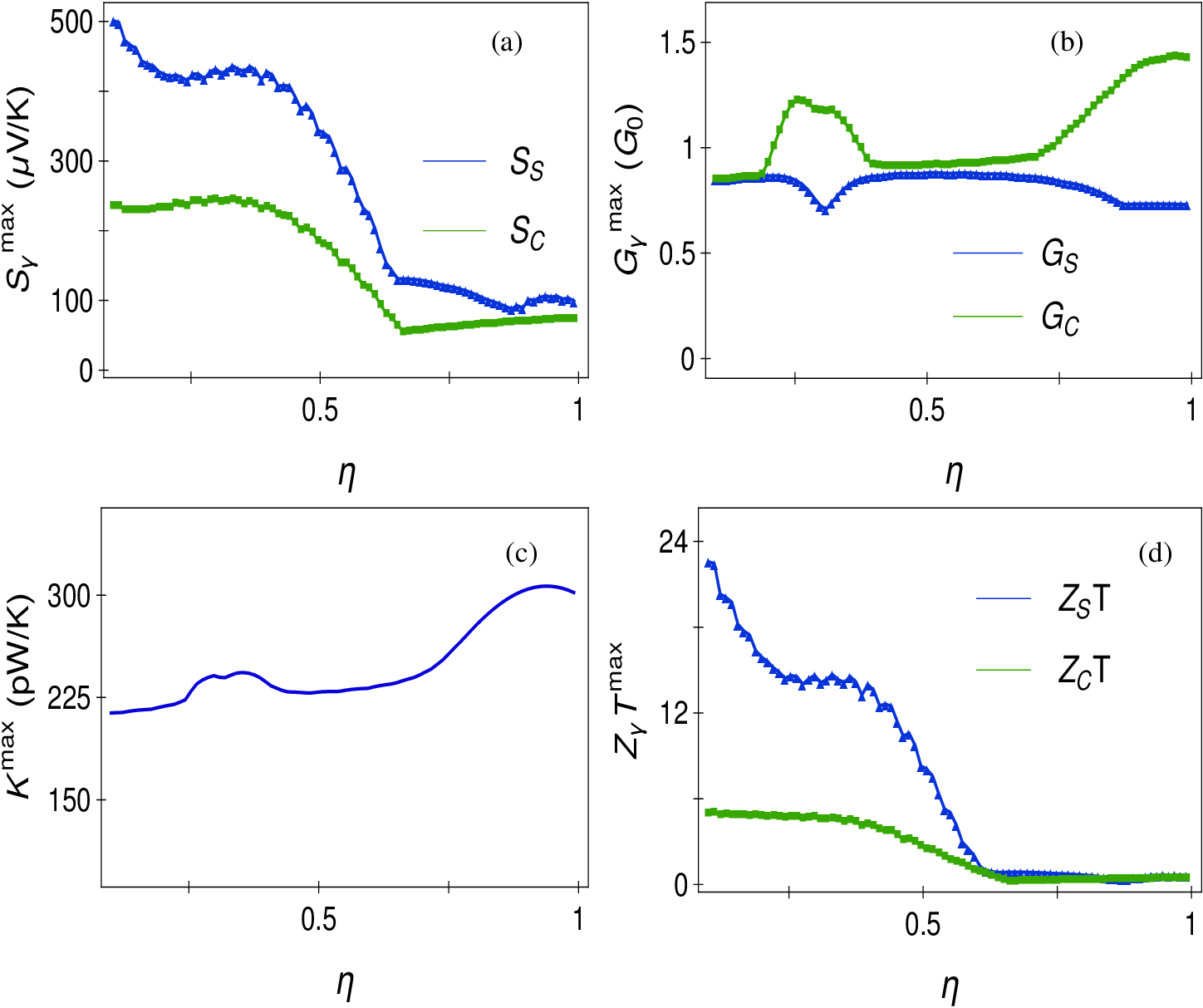}}\par}
\caption{(Color online). This plot illustrates the maximum values of thermoelectric quantities as function of $\eta$. Panel (a) shows the variation of the maximum spin and charge Seebeck coefficients, while panel (b) presents the corresponding maximum conductance values. Panel (c) depicts the variation of maximum thermal conductance for both spin and charge channels. Finally, panel (d) shows the maximum values of the spin and charge thermoelectric figure of merit ($ZT$). Together, these panels offer comprehensive insights into how each thermoelectric component evolves with the parameter $\eta$, emphasizing the tunability of thermoelectric performance under varying system conditions.}
\label{fig10}
\end{figure}

In this sub-section, we explore the impact of $\eta$, the amplitude of the cosine modulation in the next-nearest-neighbor intra-cell hopping term, on the spin and charge thermoelectric responses. The hopping between even sites is modeled as $t_N = g_b + \eta \cos(\alpha - \beta)$, allowing us to finely modulate the line shape of the transmission spectrum. By scanning over a broad range of Fermi energy, we extract the peak thermoelectric coefficients as functions of $\eta$. Interestingly, much like the role of the dimerization parameter $\delta$, $\eta$ exerts substantial control over the spin-resolved transmission profiles, effectively introducing pronounced asymmetry that directly enhances the thermoelectric figure of merit $Z_\gamma T$. The simultaneous presence of $\delta$ and $\eta$ leads to a richer parameter space, enabling more robust tuning of transport characteristics. Upon examining the variation of $Z_\gamma T$ with $\eta$, we observe consistently high values across the full range, with $Z_S T^{\text{max}}$ peaking near $\sim 22$ and $Z_C T^{\text{max}}$ around $\sim 5$. This clearly illustrates the constructive role of $\eta$ in driving thermoelectric efficiency. Complementary analysis of the Seebeck coefficients further supports this, with $S_S^{\text{max}} \sim 506$~$\mu V/K$ and $S_C^{\text{max}} \sim 240$~$\mu V/K$, reaffirming the dominance of the spin contribution. The electrical conductances exhibit a typical trend where $G_C^{\text{max}} \approx 1.5$~$G_0K$ surpasses $G_S^{\text{max}} \approx 0.9$~$G_0$, while the thermal conductance $K^{\text{max}}$ reaches a notable value of $\approx 328$~pW/K. These results are obtained under the parameter set: $\delta = 0.7$, $\beta = 0.15\pi$, $g_a = 0.4$, $g_b = 0.4$, and $\alpha = 0$. However, further exploration across other parameter regimes reveals that enhanced thermoelectric performance is not exclusive to this particular configuration. Overall, this investigation establishes $\eta$ as a potent knob for tuning the thermoelectric response, offering a clear way to achieving high $Z_\gamma T$ values through controlled modulation of transmission asymmetry.

\subsection{Dependence of Maximum Thermoelectric Parameter Values on $\alpha$}
 
\begin{figure}[ht]
{\centering \resizebox*{8cm}{8cm}{\includegraphics{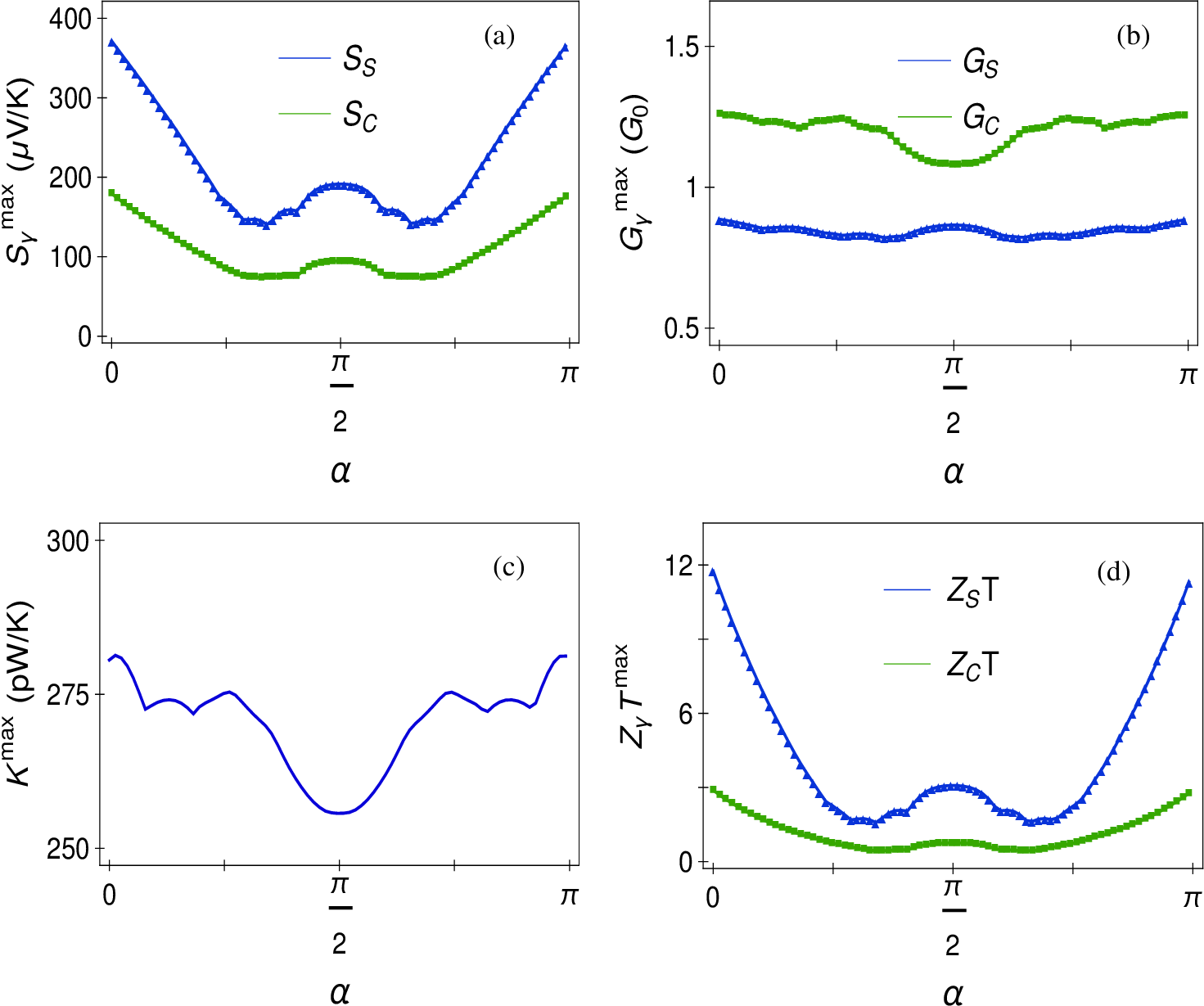}}\par}
\caption{(Color online). This plot presents the maximum values of various thermoelectric quantities as functions of the Fermi energy, evaluated across different values of the parameter $\alpha$. Panel (a) shows the variation in the peak values of spin and charge Seebeck coefficients, revealing how each spin channel contributes to the overall thermopower. Panel (b) displays the maximum spin and charge conductance, capturing the response of the system to electronic transport. Panel (c) illustrates the corresponding thermal conductance for both spin and charge carriers. Finally, panel (d) depicts the variation in the maximum thermoelectric figure of merit ($ZT$) for spin and charge channels. Together, these results provide a comprehensive understanding of how the parameter $\alpha$ modulates thermoelectric performance in the system.
}
\label{fig11}
\end{figure}

In this sub-section, we investigate how the spin and charge thermoelectric performance varies with $\alpha$, the phase angle in the cosine modulation of the next-nearest-neighbor intra-cell hopping. The hopping between even sites is modeled as $t_N = g_b + \eta \cos(\alpha - \beta)$, where $\alpha$ serves as a key tuning parameter for modulating the relative alignment of spin-resolved transmission spectra. We compute the peak thermoelectric quantities by scanning over an appropriate range of Fermi energy while varying $\alpha$ from $0$ to $\pi$, beyond which the spin channels become completely non-overlapping. Much like $\beta$, $\alpha$ plays a critical role in determining the degree of separation and overlap between up and down spin transmission functions, thereby enabling effective spin filtering. This selective overlap control directly enhances the spin Seebeck coefficient, which in turn boosts the figure of merit $Z_\gamma T$. Our analysis reveals consistently high $Z_\gamma T$ values across the entire range of $\alpha$, with $Z_S T^{\text{max}}$ reaching approximately $\sim 11$ and $Z_C T^{\text{max}}$ around $\sim 2.5$. The corresponding Seebeck coefficients exhibit peak values of $S_S^{\text{max}} \sim 370$~$\mu V/K$ and $S_C^{\text{max}} \sim 210$~$\mu V/K$, reaffirming the dominant contribution from spin transport. Additionally, the electrical conductance maintains the trend $G_C^{\text{max}} \approx 1.3~ G_0 > G_S^{\text{max}} \approx 0.9~G_0$, while the thermal conductance $K^{\text{max}}$ reaches up to $\approx 280$~pW/K. These results are obtained using $\delta = 0.5$, $\beta = 0.5\pi$, $g_a = 0.3$, $g_b = 0.3$, and $\eta = 0.3$. However, we emphasize that the emergence of high $Z_\gamma T$ values is not exclusive to this parameter set, several other configurations have also demonstrated similarly strong performance. This study confirms that the phase parameter $\alpha$ provides an effective handle for tuning spin filtering and optimizing thermoelectric response in the presence of intra-cell NNN hopping.

\subsection{\textcolor{black}{Dependence of Maximum Thermoelectric Parameter Values on $\epsilon_b$}}

\begin{figure}[ht]
	{\centering \resizebox*{8cm}{8cm}{\includegraphics{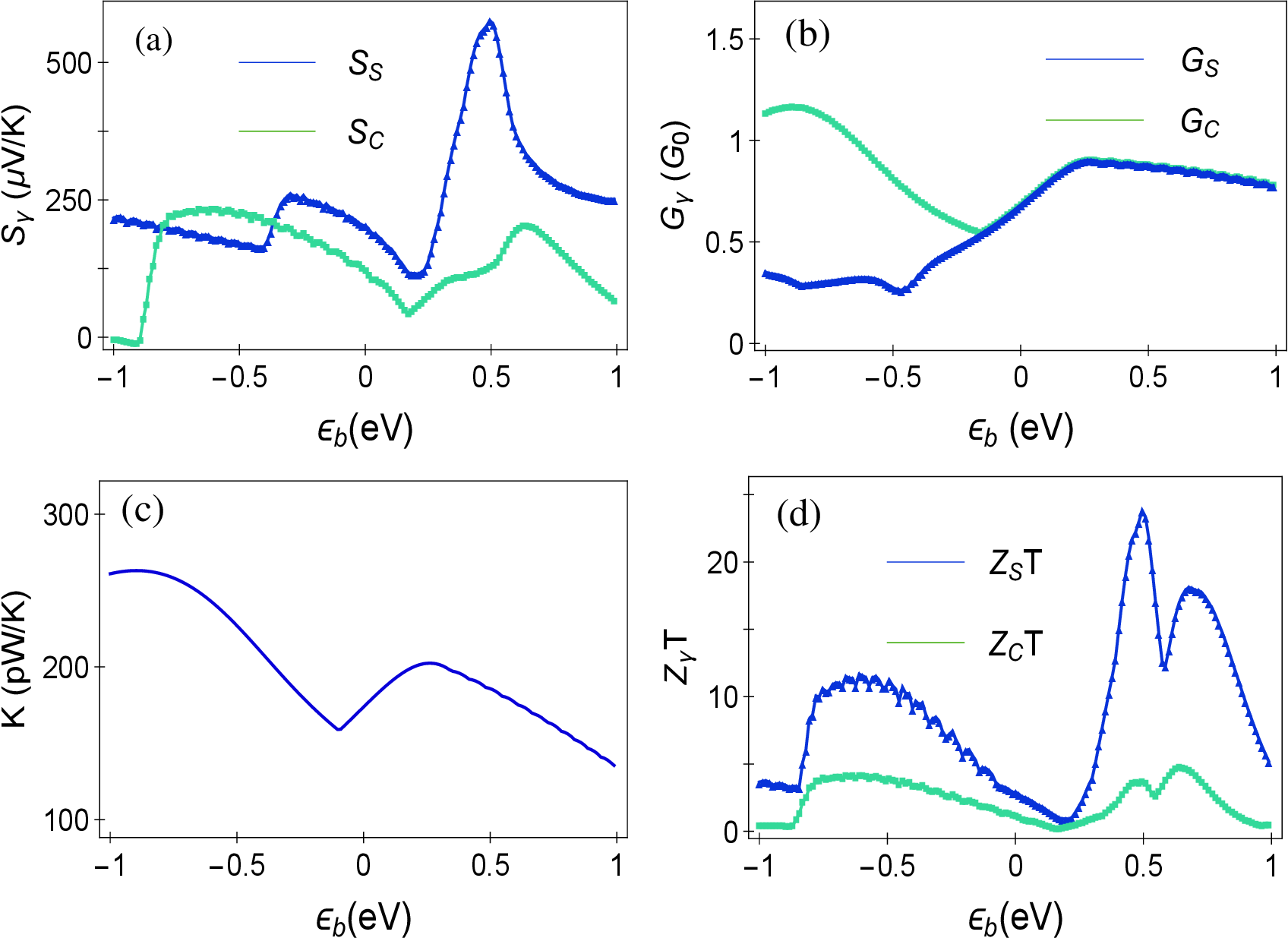}}\par}
	\caption{\textcolor{black}{(Color online). This plot presents the maximum values of various thermoelectric quantities as functions of the Fermi energy, evaluated across different values of the parameter $\epsilon_b$. Panel (a) shows the variation in the peak values of spin and charge Seebeck coefficients, revealing how each spin channel contributes to the overall thermopower. Panel (b) displays the maximum spin and charge conductance, capturing the response of the system to electronic transport. Panel (c) illustrates the corresponding thermal conductance for both spin and charge carriers. Finally, panel (d) depicts the variation in the maximum thermoelectric figure of merit ($ZT$) for spin and charge channels.}
			}
			\label{fig12}
		\end{figure}
		
\textcolor{black}{In this sub-section, we investigate how the spin and charge thermoelectric performance varies with onsite energy energy differences, here we fix $\epsilon_a = 0.5$~eV and varied $\epsilon_b$ from $-1$ to $+1$~eV. We compute the peak thermoelectric quantities by scanning over an appropriate range of Fermi energy. Our analysis reveals consistently high $Z_\gamma T$ values across the entire range of $\alpha$, with $Z_S T^{\text{max}}$ reaching approximately $\sim 25$ and $Z_C T^{\text{max}}$ around $\sim 4$. The corresponding Seebeck coefficients exhibit peak values of $S_S^{\text{max}} \sim 540$~$\mu V/K$ and $S_C^{\text{max}} \sim 250$~$\mu V/K$, reaffirming the dominant contribution from spin transport. Additionally, the electrical conductance maintains the trend $G_C^{\text{max}} \approx 1.2~G_0 > G_S^{\text{max}} \approx 0.9~G_0$, while the thermal conductance $K^{\text{max}}$ reaches up to $\approx 280$~pW/K. These results are obtained using $\delta = 0.8$, $\beta = 0.7\pi$, $g_a = 0.5$, $g_b = 0.5$, and $\eta = 0.5$, $\alpha = 0.75\pi$. However, we emphasize that the emergence of high $Z_\gamma T$ values is not exclusive to this parameter set, several other configurations have also demonstrated similarly strong performance. Here the onsite energy differences in the Fibonacci site enhances disorder into the system but since it is deterministic type of disorder and the system size is small the system does not localizes but only the transmission line shape becomes asymmetric. As a result of which $Z_\gamma T$ enhances.  This study confirms that the difference in onsite energies of the Fibonacci chain provides an effective handle for tuning spin filtering and optimizing thermoelectric response in the presence of intra-cell NNN hopping.}

\subsection{Dependence of Maximum Thermoelectric Quantities on $\delta$ in a Higher-Order Fibonacci Chain}

\begin{figure}[ht]
	{\centering \resizebox*{8cm}{8cm}{\includegraphics{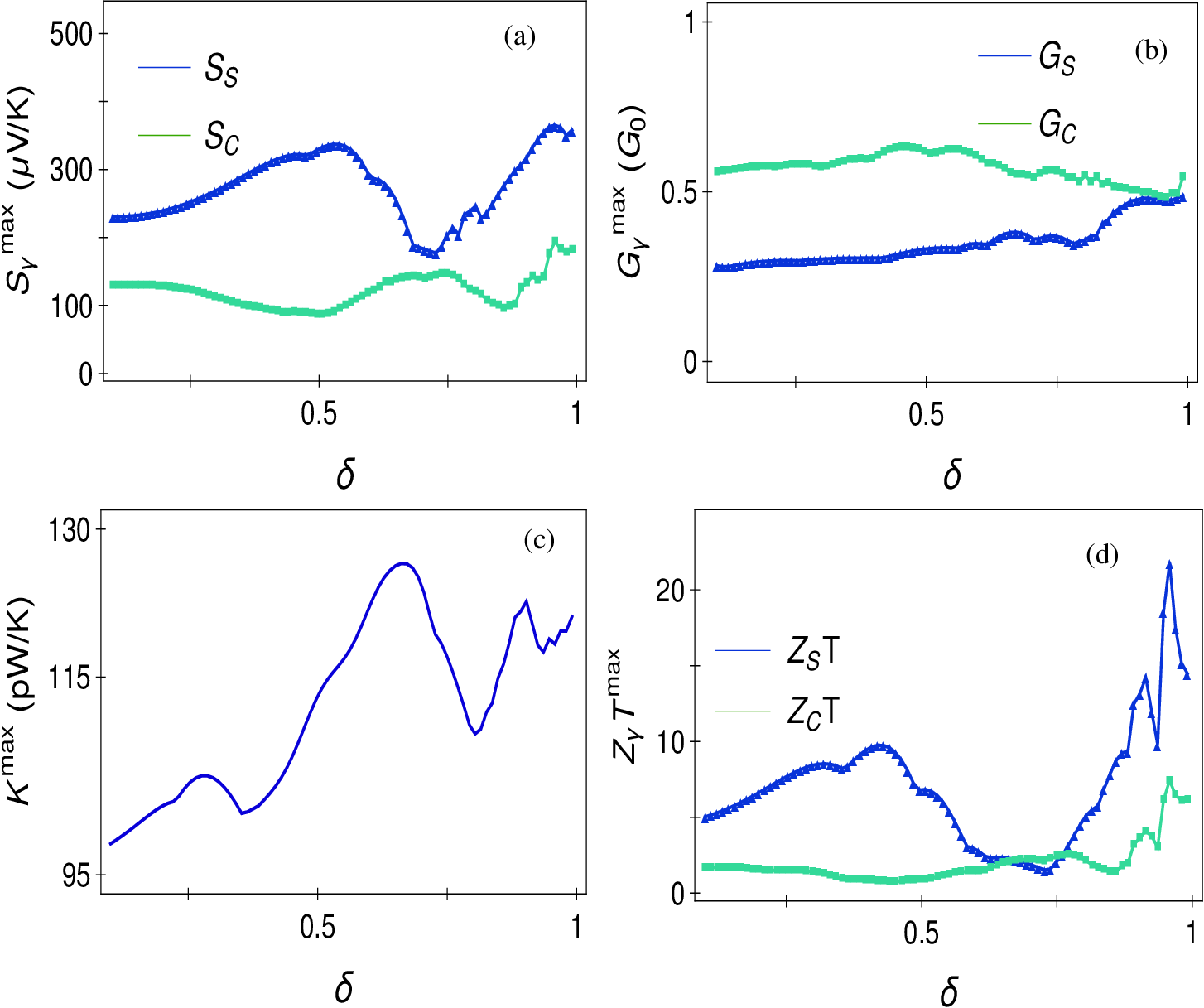}}\par}
	\caption{(Color online). Maximum values of various thermoelectric quantities as functions of the parameter $\delta$. (a) Maximum spin and charge Seebeck coefficients; (b) maximum spin and charge electrical conductances; (c) maximum thermal conductance; and (d) maximum thermoelectric figure of merit ($ZT$) for spin and charge channels. The plots reveal the evolution of each thermoelectric component with $\delta$, offering insight into the system’s response. Calculations are performed for a larger Fibonacci generation with a total of 34 sites.}
	\label{fig13}
\end{figure}

In this subsection, we investigate the effect of the modulation strength $\delta$, which governs the amplitude of the cosine-modulated nearest-neighbor hopping, on the spin and charge thermoelectric (TE) transport properties of a higher-order Fibonacci chain. Specifically, we focus on the $8^{\text{th}}$ generation Fibonacci sequence, comprising a total of 34 lattice sites. Examining such a larger system size is essential to assess the scalability and practical viability of the model as a potential thermoelectric generator. To quantitatively evaluate the system’s thermoelectric response, we compute the maximum values of relevant thermoelectric quantities by performing a comprehensive scan over a wide range of Fermi energies for each value of $\delta$. This approach allows us to analyze how the optimal performance of the system evolves with increasing hopping modulation. The underlying motivation is to determine whether the system can consistently deliver high spin and charge figure-of-merit values, $Z_\gamma T$ ($\gamma = S, C$), across a broad parameter regime, or whether its thermoelectric efficiency is restricted to narrowly tuned conditions. Such an evaluation is crucial for assessing the robustness and reliability of the model as a practical TE platform. Our numerical results demonstrate that the maximum values of $Z_\gamma T$ remain significantly large over the entire range of $\delta$ considered, with values typically exceeding 3, thereby affirming the system’s stable and efficient performance. In particular, the spin figure of merit, $Z_S T^{\text{max}}$, exhibits a striking peak of approximately $\sim 21$, while the charge counterpart, $Z_C T^{\text{max}}$, reaches a value around $\sim 4$. These results underscore the dominant contribution of the spin channel to the overall TE response. To gain deeper insight into the origin of this enhancement, we analyze the corresponding Seebeck coefficients. We find that the maximum spin Seebeck coefficient reaches $S_S^{\text{max}} \sim 320\,\mu$V/K, whereas the maximum charge Seebeck coefficient attains $S_C^{\text{max}} \sim 170\,\mu$V/K. These observations further reinforce the spin-selective nature of the thermoelectric response.
Additionally, the electrical conductance results show that $G_C^{\text{max}} \approx 0.7\,G_0$ and $G_S^{\text{max}} \approx 0.5\,G_0$, where $G_0$ is the conductance quantum, indicating a reasonably good transport regime. The maximum thermal conductance is found to be $K^{\text{max}} \approx 125$~pW/K. The simulations are carried out using the following set of parameters: $\beta = 0.3\pi$, $g_a = 0.3$, $g_b = 0.3$, $\eta = 0.5$, and $\alpha = 0.2\pi$. It is worth noting that the enhanced values of $Z_\gamma T$ reported here are not unique to this specific parameter set; comparable results are obtained for other configurations as well, demonstrating the generality of the mechanism. Interestingly, despite the Seebeck coefficients in this larger system being relatively lower than those observed in earlier cases, the spin figure of merit remains significantly high. This enhancement can be attributed to the increased system size, which results in reduced electronic thermal conductance, appearing in the denominator of the expression for $Z_\gamma T$ [see Eq.~\eqref{eq19}], and thereby leads to a higher overall value of the spin figure of merit. In summary, the observed robustness of $Z_\gamma T$ with respect to variations in the hopping modulation strength $\delta$ highlights the effectiveness of the model in sustaining high thermoelectric efficiency over a wide parameter space. These findings support the suitability of the Fibonacci-modulated chain as a promising candidate for next-generation thermoelectric applications.

\textcolor{black}{\textbf{Experimental setup:} The SSH model has been realized in laboratories using various approaches. One of the most prominent platforms is cold atoms trapped in optical lattices or superlattices, where a one-dimensional optical superlattice with alternating well depths or spacings naturally leads to alternating tunneling amplitudes between neighboring sites~\cite{essh}. In such systems, the hopping parameters can be finely tuned by adjusting the lattice depth, the relative offset of the double-well superlattice, or the physical separation between wells. Since tunneling amplitudes depend exponentially on the barrier height and width, even modest modifications enable a wide dynamic range for controlling intra- and inter-cell hoppings. Furthermore, several schemes have been proposed to extend this control to next-nearest-neighbor (NNN) hopping terms~\cite{essh1}. With these experimental capabilities, we believe that our proposed method can be effectively implemented in the laboratory to test for enhanced spin thermoelectric efficiency.  }

\subsection{Justification of our model}

\textcolor{black}{We study a one-dimensional SSH chain that includes both nearest-neighbor (NN) and next-nearest-neighbor (NNN) intra- and inter-cell hoppings. A cosine modulation is introduced to fine-tune the hopping amplitudes in a controlled manner. Hopping modulation has previously been studied mainly through disorder-based approaches. For instance, random disorder eliminates extended states altogether, leading to Anderson localization~\cite{ander}. In contrast, the Aubry–André–Harper (AAH)~\cite{aah1,aah2,aah3,jap} model, depending on parameter choices, which interpolates between periodic and quasi-periodic regimes, giving rise to regimes where localized and extended states coexist. Our approach differs in that it does not rely on any form of disorder. Instead, the cosine term provides a uniform, site-independent modulation of hopping strengths. This simple yet flexible scheme enables precise control of the electronic structure and results in strong modifications of the transport properties, most notably a pronounced separation between spin channels in the transmission spectrum.}   

\begin{table*}[ht]
\centering
\renewcommand{\arraystretch}{1.3}
\small 
\begin{tabular}{|c|c|c|}
\hline
\multicolumn{1}{|c|}{\textbf{Parameters}} & \multicolumn{1}{|c|}{\textbf{Spin TE parameters}} & \multicolumn{1}{|c|}{\textbf{Charge TE parameters}}  \\
\hline
\begin{tabular}[c]{@{}l@{}}
\hspace{1.5cm} \textbf{$\delta$}\\
 modulation strength of \\ the cosine term in NNH \\ $(t_{1(2)} = t[1 \pm \delta \cos(\beta)])$ 
\end{tabular}  & 
\begin{tabular}[c]{@{}l@{}}
$\bullet$ $Z_{S}T^{max}$  $\approx18$\\
$\bullet$ $S_{S}^{max}$ $\approx470$~$\mu$V/K\\
$\bullet$ $G_S^{max}$  $\approx0.6~G_0$
\end{tabular} &
\begin{tabular}[c]{@{}l@{}}
$\bullet$ $Z_{C}T^{max}$  $\approx4$\\
$\bullet$ $S_C^{max}$  $\approx270$~$\mu$V/K\\
$\bullet$ $G_C^{max}$  $\approx1.4~G_0$
\end{tabular} \\
\hline
\begin{tabular}[c]{@{}l@{}}
\hspace{1.5cm} \textbf{$\beta$}\\
the phase angle  of  \\ cosine term in NNH 
\end{tabular} &
\begin{tabular}[c]{@{}l@{}}
$\bullet$ $Z_{S}T^{max}$  $\approx21$\\
$\bullet$ $S_S^{max}$  $\approx500$~$\mu$V/K\\
$\bullet$ $G_S^{max}$  $\approx1~G_0$
\end{tabular} &
\begin{tabular}[c]{@{}l@{}}
$\bullet$ $Z_{C}T^{max}$  $\approx5$\\
$\bullet$ $S_C^{max}$  $\approx270$~$\mu$V/K\\
$\bullet$ $G_C^{max}$  $\approx1.5~G_0$
\end{tabular} \\
\hline
\begin{tabular}[c]{@{}l@{}}
\hspace{1.5cm} \textbf{$g_a$}\\
first term in NNNH for odd sites \\ $(t_{M(N)} = g_{a(b)} + \eta \cos(\alpha \pm \beta)])$ 
\end{tabular} & 
\begin{tabular}[c]{@{}l@{}}
$\bullet$ $Z_{S}T^{max}$  $\approx18$\\
$\bullet$ $S_S^{max}$  $\approx470$~$\mu$V/K\\
$\bullet$ $G_S^{max}$  $\approx1~G_0$
\end{tabular} &
\begin{tabular}[c]{@{}l@{}}
$\bullet$ $Z_{C}T^{max}$ $\approx3$\\
$\bullet$ $S_C^{max}$  $\approx240$~$\mu$V/K\\
$\bullet$ $G_C^{max}$  $\approx1.6~G_0$
\end{tabular} \\
\hline
\begin{tabular}[c]{@{}l@{}}
\hspace{1.5cm} \textbf{$g_b$}\\
first term in NNNH \\ for even sites
\end{tabular} & 
\begin{tabular}[c]{@{}l@{}}
$\bullet$ $Z_{S}T^{max}$  $\approx29$\\
$\bullet$ $S_S^{max}$  $\approx470$~$\mu$V/K\\
$\bullet$ $G_S^{max}$  $\approx1~G_0$
\end{tabular} &
\begin{tabular}[c]{@{}l@{}}
$\bullet$ $Z_{C}T^{max}$  $\approx4$\\
$\bullet$ $S_C^{max}$  $\approx240$~$\mu$V/K\\
$\bullet$ $G_C^{max}$  $\approx1.6~G_0$
\end{tabular} \\
\hline
\begin{tabular}[c]{@{}l@{}}
\hspace{1.5cm} \textbf{$\eta$}\\
modulation strength \\ of  the cosine term in NNNH 
\end{tabular} & 
\begin{tabular}[c]{@{}l@{}}
$\bullet$ $Z_{S}T^{max}$  $\approx22$\\
$\bullet$ $S_S^{max}$  $\approx500$~$\mu$V/K\\
$\bullet$ $G_S^{max}$ $\approx0.9~G_0$
\end{tabular} &
\begin{tabular}[c]{@{}l@{}}
$\bullet$ $Z_{C}T^{max}$  $\approx5$\\
$\bullet$ $S_C^{max}$  $\approx240$~$\mu$V/K\\
$\bullet$ $G_C^{max}$  $\approx1.5~G_0$
\end{tabular} \\
\hline
\begin{tabular}[c]{@{}l@{}}
\hspace{1.5cm} \textbf{$\alpha$}\\
phase angle  of the cosine \\ term in NNNH 
\end{tabular} & 
\begin{tabular}[c]{@{}l@{}}
$\bullet$ $Z_{S}T^{max}$  $\approx11$\\
$\bullet$ $S_S^{max}$ $\approx370$~$\mu$V/K\\
$\bullet$ $G_S^{max}$  $\approx0.9~G_0$
\end{tabular} &
\begin{tabular}[c]{@{}l@{}}
$\bullet$ $Z_{C}T^{max}$  $\approx2.5$\\
$\bullet$ $S_C^{max}$ $\approx200$~$\mu$V/K\\
$\bullet$ $G_C^{max}$  $\approx1.3~G_0$
\end{tabular} \\
\hline
\end{tabular}
\caption{Variation of spin and charge TE properties with adjustable nearest- and next-nearest-neighbor hopping configurations}
\label{tab:model_comparison}
\end{table*}

\section{Conclusion}
\label{sec:conclusion}

To conclude, our study provides a comprehensive \textcolor{black}{qualitative} exploration of the thermoelectric properties of an SSH chain incorporating both nearest-neighbor (NN) and next-nearest-neighbor (NNN) hopping. These hopping interactions modify the line shape and position of the up-spin and down-spin transmission spectra, thereby effectively controlling the separation of spin channels which is a key factor governing spin thermoelectric efficiency. By systematically analyzing the effects of these hopping integrals along with correlated diagonal disorder in the form of a Fibonacci potential, we reveal significant enhancements in the thermoelectric figure of merit ($ZT$). The interplay between NN and NNN not only improves the efficiency of spin-dependent thermoelectricity but also enables tunability of the overall thermoelectric performance, paving the way toward high-efficiency operation. These results offer valuable insights into the capabilities of nanoscale molecular systems for next-generation energy conversion technologies, underscoring their potential for the design of advanced thermoelectric devices.

$\bullet$ \textbf{Efficient spin channel separation:} The interplay of NN and NNN hopping strengths serves as a powerful mechanism for tailoring spin-dependent transmission, enabling selective control of up- and down-spin channels.

$\bullet$ \textbf{High-Performance Spin Thermoelectric Response:} Spin-mediated thermoelectric transport exhibits higher efficiency than conventional charge-based transport. The inherently non-dissipative character of spin currents allows the spin Seebeck effect to deliver markedly improved thermoelectric performance.

$\bullet$ \textbf{Impact of Fibonacci Pattern:} The introduction of diagonal site potentials arranged in a Fibonacci pattern induces marked changes in the localization–delocalization transition, thereby enhancing the thermoelectric figure of merit ($ZT$) for both spin and charge transport.

$\bullet$ \textbf{Improved $ZT$ Metrics:} The results indicate that strategic interplay between hopping parameters and Fibonacci pattern enables a significant increase in the thermoelectric figure of merit $ZT$ for spin and charge channels, reinforcing the prospects for high-performance thermoelectric technologies.

$\bullet$ \textbf{Framework for Efficient Thermoelectric Devices:} Our study highlights fundamental strategies for engineering efficient thermoelectric devices, identifying nanoscale molecular architectures as promising platforms for future energy conversion.

$\bullet$ \textbf{Emerging Insights into Nanoscale Thermoelectricity:} These findings open new avenues for tailoring thermoelectric performance at the molecular level, underscoring their potential in next-generation nanoscale energy conversion technologies.

\appendix
\section{}

\subsection{Characterization of the Maximum Thermoelectric Parameter Variations as Functions of $\alpha$ and $\beta$}

\begin{figure*}[ht]
	{\centering \resizebox*{16.0cm}{8cm}{\includegraphics{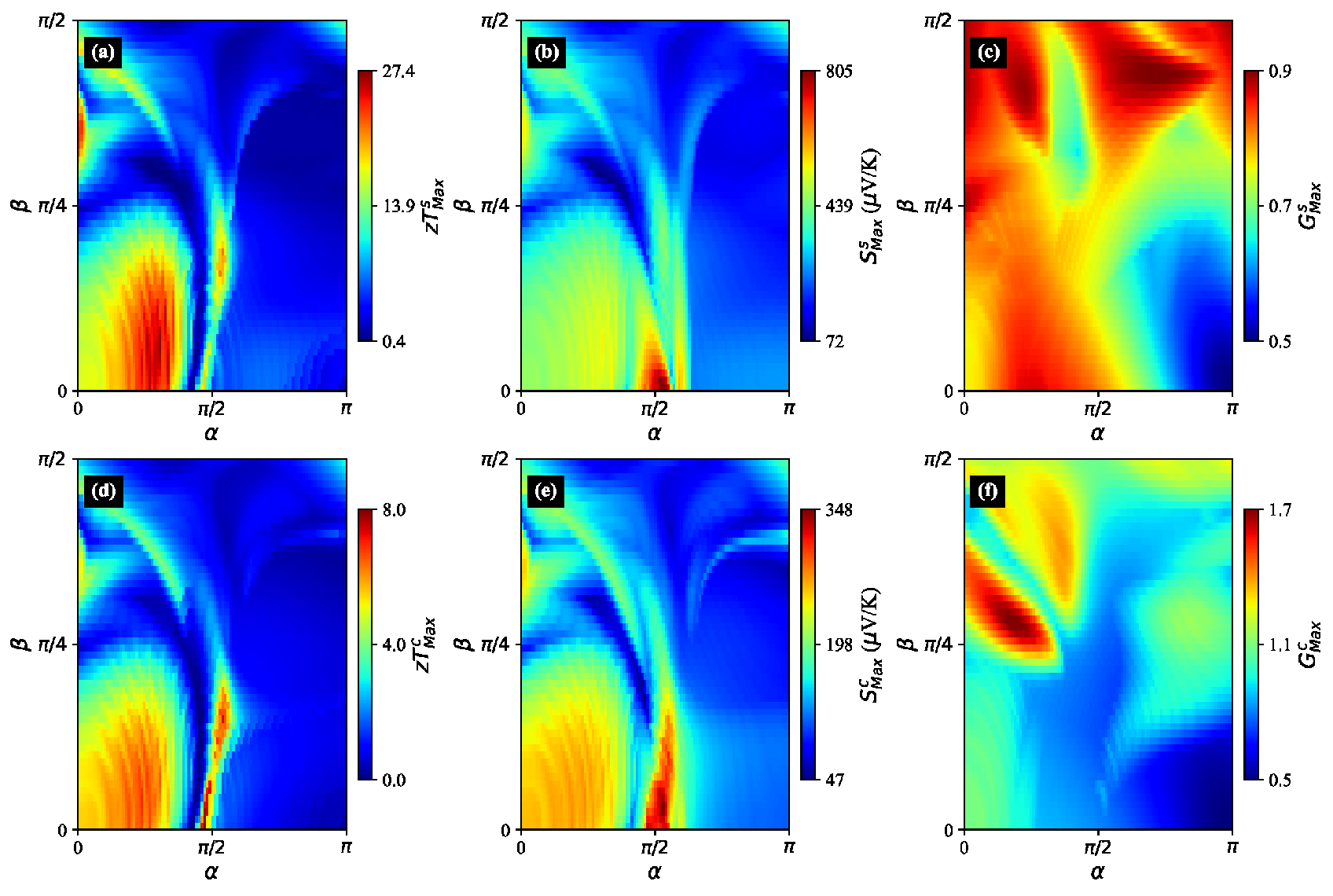}}\par}
\caption{(Color online). This figure presents the phase diagram of thermoelectric parameters as functions of both $\alpha$ and $\beta$. Panel (a) and (d) display the variation of the maximum values of the spin and charge thermoelectric figure of merit ($ZT$), respectively. Panels (b) and (e) illustrate the corresponding maximum values of the spin and charge Seebeck coefficients. Finally, panels (c) and (f) show the maximum spin and charge conductance. Together, these panels provide a comprehensive overview of how the thermoelectric response for both spin and charge channels is modulated by the interplay between $\alpha$ and $\beta$.}
			\label{fig14}
		\end{figure*}

\begin{figure*}[ht]
			{\centering \resizebox*{16.0cm}{8cm}{\includegraphics{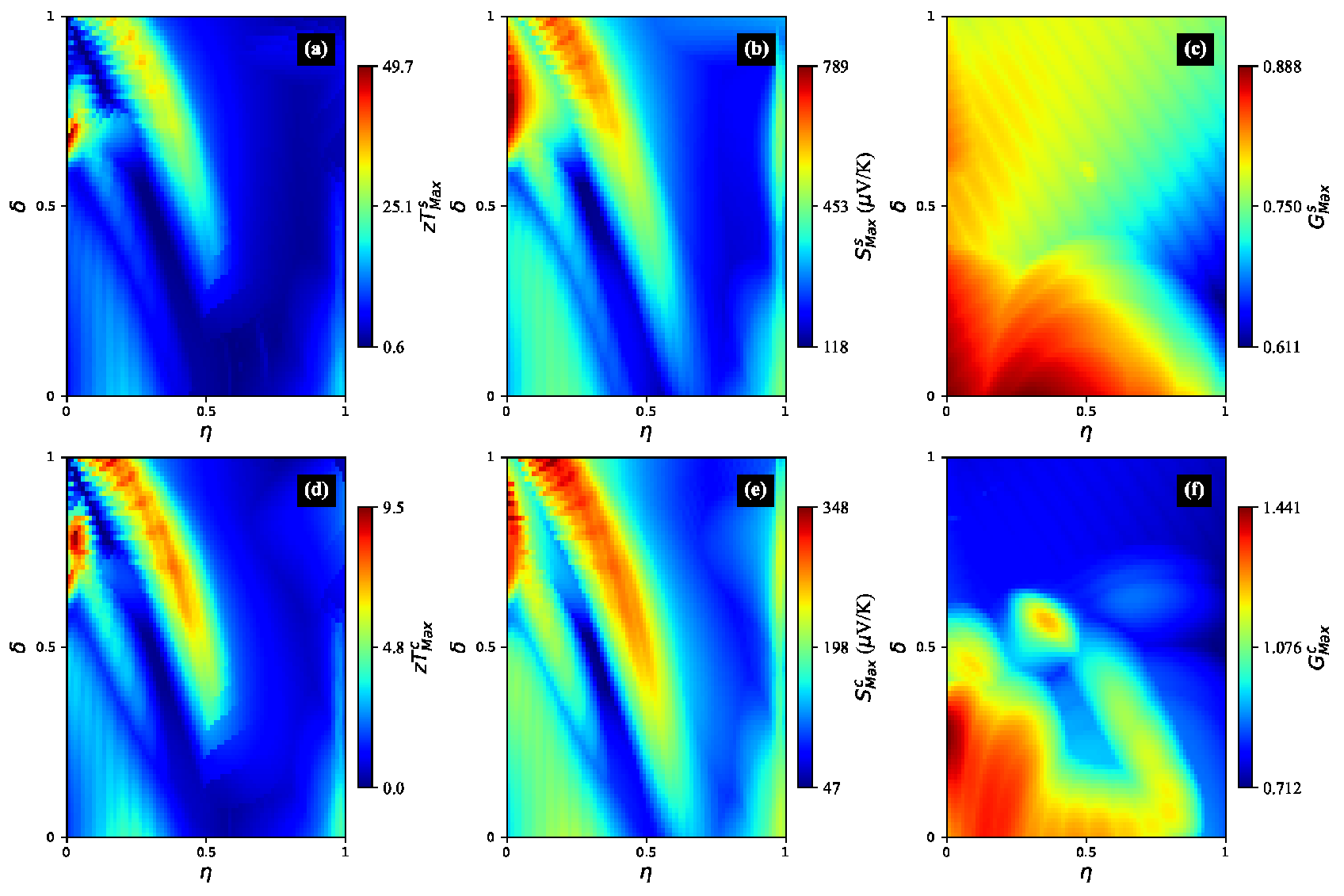}}\par}
\caption{(Color online). This figure illustrates the phase diagrams of thermoelectric parameters as functions of both $\delta$ and $\eta$. Panels (a) and (d) depict the variation of the maximum values of the spin and charge thermoelectric figure of merit ($ZT$), respectively. Panels (b) and (e) show the corresponding maximum values of the spin and charge Seebeck coefficients, while panels (c) and (f) present the maximum values of spin and charge conductance. Collectively, these results provide a comprehensive understanding of how thermoelectric properties in both spin and charge channels evolve with the parameters $\delta$ and $\eta$.}
					\label{fig15}
				\end{figure*}

In this sub-section, we explore the combined influence of $\alpha$ and $\beta$, the phase angles associated with the cosine modulations in next-nearest-neighbor and nearest-neighbor hopping, respectively, on spin and charge thermoelectric performance. We construct density plots of the peak values of the thermoelectric coefficients by scanning over a broad range of Fermi energies and analyzing their variation with respect to $\alpha$ and $\beta$. As both parameters primarily modulate the positions of the spin-resolved transmission spectra, their simultaneous tuning offers enhanced control over the overlap and separation between up and down spin channels. This, in turn, can strongly influence thermoelectric efficiency. Our results indicate that the maximum values reach $Z_S T^{\text{max}} \sim 27$ and $Z_C T^{\text{max}} \sim 8$, demonstrating substantial enhancement. The peak Seebeck coefficients attain values as high as $S_S^{\text{max}} \sim 800~\mu V/K$ and $S_C^{\text{max}} \sim 340~\mu V/K$, highlighting the significant role of spin-dependent filtering. As observed in previous cases, the spin channel again dominates with $S_S^{\text{max}} > S_C^{\text{max}}$. The electrical conductances follow the expected hierarchy, with $G_C^{\text{max}} \approx 1.7~G_0$ exceeding $G_S^{\text{max}} \approx 0.9~G_0$. These findings confirm that concurrent variation of both phase angles, $\alpha$ and $\beta$, enables fine-tuned control of transmission characteristics, yielding remarkably high figures of merit $Z_\gamma T$ and establishing this model as a highly efficient spin-dependent thermoelectric platform.

\begin{figure*}[ht]
{\centering \resizebox*{16.0cm}{7.9cm}{\includegraphics{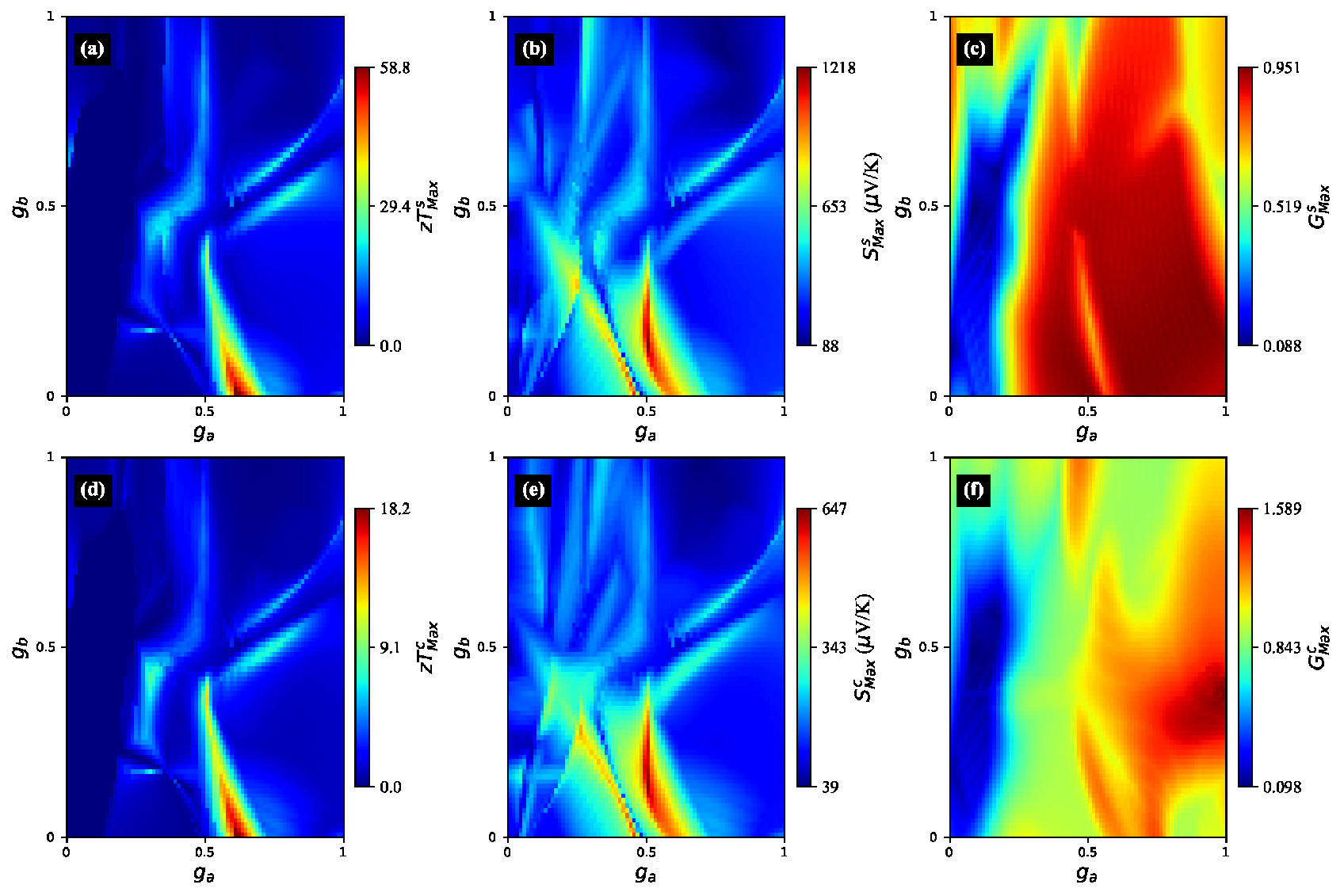}}\par}
\caption{(Color online).  This figure presents phase diagrams of thermoelectric parameters as functions of both $g_a$ and $g_b$. Panels (a) and (d) show the variation of the maximum values of the spin and charge thermoelectric figure of merit ($ZT$), respectively. Panels (b) and (e) display the corresponding maximum values of the spin and charge Seebeck coefficients, while panels (c) and (f) illustrate the variation in the maximum spin and charge conductance. Together, these results provide detailed insights into how the thermoelectric performance in spin and charge channels evolves with $g_a$ and $g_b$.	}
	\label{fig16}
	\end{figure*}
\subsection{Maximum Thermoelectric Parameter Variations as a Function of $\delta$ and $\eta$}
				
In this sub-section, we examine the combined effect of $\eta$ and $\delta$, the amplitude parameters associated with the cosine modulations in next-nearest-neighbor and nearest-neighbor hopping, respectively, on the charge and spin thermoelectric coefficients. We generate density plots of the peak thermoelectric quantities by scanning over a broad range of Fermi energies and analyzing their variation as functions of $\eta$ and $\delta$. Since both parameters significantly influence the shape and position of the spin-resolved transmission functions, their simultaneous tuning enables enhanced control over transmission asymmetry and overlap. Our results reveal that the maximum figures of merit reach $Z_S  T^{\text{max}} \sim 49$ and $Z_C T^{\text{max}} \sim 9$, indicating remarkable thermoelectric efficiency in the spin channel. The Seebeck coefficients also exhibit significant enhancement, with $S_S^{\text{max}} \sim 780~\mu V/K$ and $S_C^{\text{max}} \sim 340~\mu V/K$, further reinforcing the advantage of spin-based transport. As in previous observations, $S_S^{\text{max}}$ exceeds $S_C^{\text{max}}$, confirming superior spin thermopower. The charge conductance maintains its dominance over the spin counterpart, with $G_C^{\text{max}} \approx 1.4~G_0$ and $G_S^{\text{max}} \approx 0.8~G_0$. These findings demonstrate that  simultaneous modulation of $\delta$ and $\eta$ offers a robust tuning mechanism to achieve high-performance thermoelectric response in both charge and spin channels.
				
\subsection{Dependence of Maximum Values of Thermoelectric Parameters on $g_a$ and $g_b$}
				
In this sub-section, we investigate the influence of $g_a$ and $g_b$, the amplitude parameters associated with the cosine modulations in the nearest-neighbor and next-nearest-neighbor hopping terms, respectively, on spin and charge thermoelectric performance. To this end, we construct density plots of the peak thermoelectric parameters by scanning over a sufficiently wide range of Fermi energies and examining their dependence on $g_a$ and $g_b$. As both $g_a$ and $g_b$ modulate the profile and position of the transmission functions, their simultaneous tuning allows for refined control over spin-dependent transport characteristics. Our results demonstrate a substantial enhancement in thermoelectric efficiency, with the maximum figures of merit reaching $Z_{S}^{\text{max}} T \sim 58$ and $Z_{C}^{\text{max}} T \sim 18$. The corresponding spin and charge Seebeck coefficients attain high values, $S_{S}^{\text{max}} \sim 1200~\mu V/K$ and $S_{C}^{\text{max}} \sim 640~\mu V/K$, underscoring the critical role of thermopower in boosting overall thermoelectric performance. As previously observed, the spin Seebeck coefficient dominates its charge counterpart. We also analyze the variation in electronic conductance and find that $G_{C}^{\text{max}} \approx 1.5~G_0$ remains higher than $G_{S}^{\text{max}} \approx 0.9~G_0$. These findings confirm that the simultaneous tuning of $g_a$ and $g_b$ serves as an effective strategy for achieving high values of $Z_\gamma T$ and optimizing both spin and charge thermoelectric properties.

\section{}
\subsection{\textcolor{black}{Dependence of Maximum Thermoelectric Quantities on $\delta$ in a Higher-Order Fibonacci Chain}}

\begin{figure}[ht]
	{\centering \resizebox*{8cm}{8cm}{\includegraphics{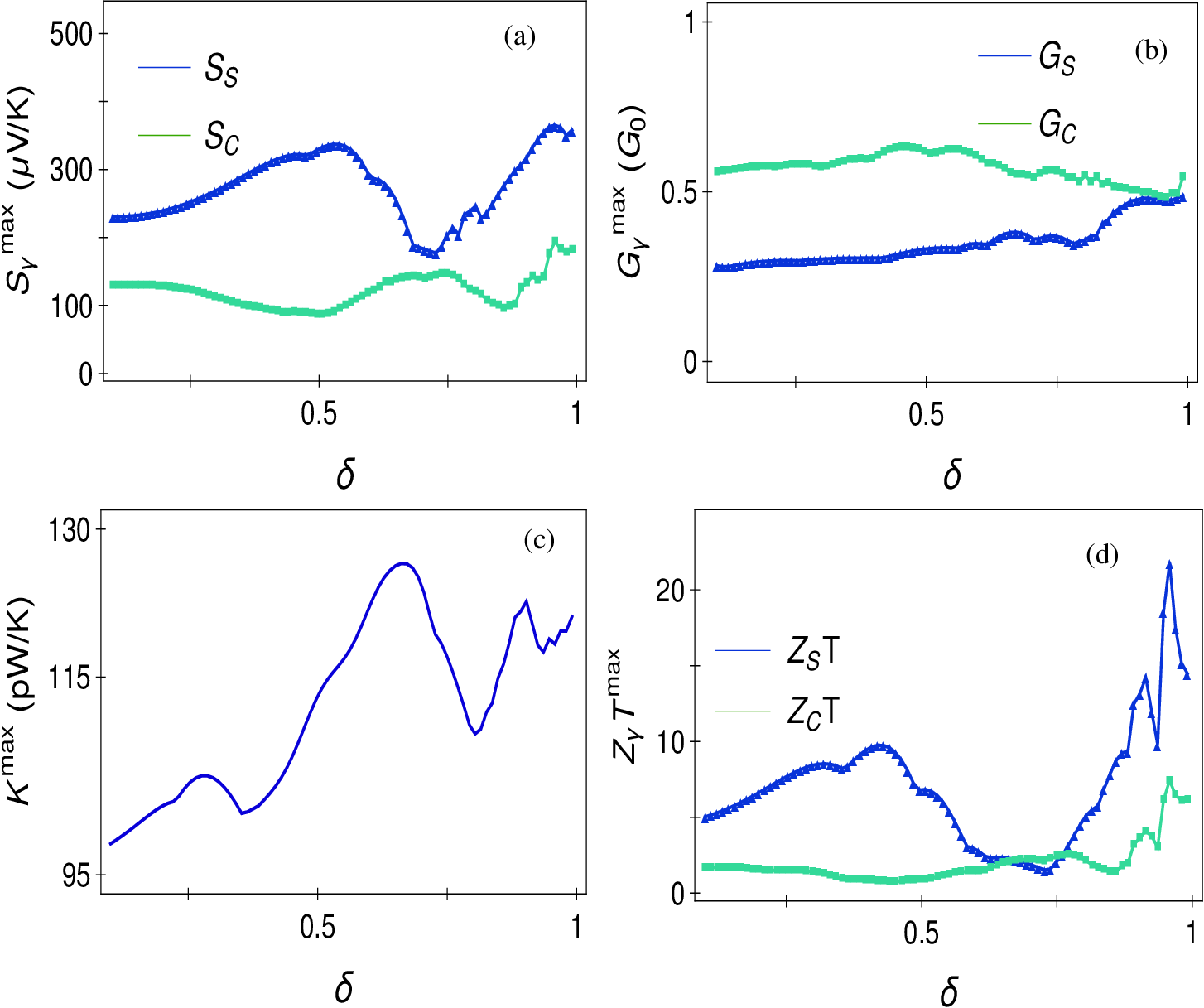}}\par}
	\caption{\textcolor{black}{(Color online). Maximum values of various thermoelectric quantities as functions of the parameter $\delta$. (a) Maximum spin and charge Seebeck coefficients; (b) maximum spin and charge electrical conductances; (c) maximum thermal conductance; and (d) maximum thermoelectric figure of merit ($ZT$) for spin and charge channels. The plots reveal the evolution of each thermoelectric component with $\delta$, offering insight into the system’s response. Calculations are performed for a larger Fibonacci generation with a total of $144$ sites.}}
			\label{fig17}
		\end{figure}

\textcolor{black}{In this subsection, we investigate the effect of the modulation strength $\delta$, which governs the amplitude of the cosine-modulated nearest-neighbor hopping, on the spin and charge thermoelectric (TE) transport properties of a higher-order Fibonacci chain. Specifically, we focus on the $11^{\text{th}}$ generation Fibonacci sequence, comprising a total of $144$ lattice sites to check how this model behave in larger system sizes. Computing the maximum values of relevant thermoelectric quantities, scanning over a wide range of Fermi energies for each value of $\delta$. Our numerical results demonstrate that the maximum values of $Z_\gamma T$ remain significantly large over the entire range of $\delta$ considered, thereby affirming the system’s stable and efficient performance. In particular, the spin figure of merit, $Z_S T^{\text{max}}$, exhibits a striking peak of approximately $\sim 23$, while the charge counterpart, $Z_C T^{\text{max}}$, reaches a value around $\sim 5$. These results underscore the dominant contribution of the spin channel to the overall TE response. To gain deeper insight into the origin of this enhancement, we analyze the corresponding Seebeck coefficients. We find that the maximum spin Seebeck coefficient reaches $S_S^{\text{max}} \sim 400\,\mu$V/K, whereas the maximum charge Seebeck coefficient attains $S_C^{\text{max}} \sim 200\,\mu$V/K. These observations further reinforce the spin-selective nature of the thermoelectric response. Additionally, the electrical conductance results show that $G_C^{\text{max}} \approx 0.7\,G_0$ and $G_S^{\text{max}} \approx 0.5\,G_0$,  The maximum thermal conductance is found to be $K^{\text{max}} \approx 125$~pW/K. The simulations are carried out using same set of parameters as with order number $8$.}

\end{document}